\documentclass[10pt]{article}
\usepackage{graphicx,xcolor,amsmath,amssymb,multirow,ulem,wrapfig,amsfonts,natbib}
\usepackage{epstopdf,epsfig,tikz,amsbsy,mathabx,esint,longtable,subfigure,subcaption,placeins}
\usetikzlibrary{arrows,decorations.markings}
\usetikzlibrary{patterns,positioning,decorations.pathmorphing,arrows,decorations.pathreplacing,calligraphy}
 
\definecolor{OliveGreen}{rgb}{0,0.6,0}

\renewcommand{\Re}{\mathrm{Re}}
\pgfdeclarelayer{back} \pgfsetlayers{back,main}
\definecolor{OliveGreen}{rgb}{0,0.6,0}

\def\XXint#1#2#3{{\setbox0=\hbox{$#1{#2#3}{\int}$}
		\vcenter{\hbox{$#2#3$}}\kern-.5\wd0}}

\makeatletter
\pgfkeys{%
	/tikz/on layer/.code={
		\pgfonlayer{#1}\begingroup
		\aftergroup\endpgfonlayer
		\aftergroup\endgroup
	},
	/tikz/node on layer/.code={
		\pgfonlayer{#1}\begingroup
		\expandafter\def\expandafter\tikz@node@finish\expandafter{\expandafter\endgroup\expandafter\endpgfonlayer\tikz@node@finish}%
	},
}

\newcommand{\tstar}[5]{% inner radius, outer radius, tips, rot angle, options
	\pgfmathsetmacro{\starangle}{360/#3}
	\draw[#5] (#4:#1)
	\foreach \x in {1,...,#3}
	{ -- (#4+\x*\starangle-\starangle/2:#2) -- (#4+\x*\starangle:#1)
	}
	-- cycle;
}

\usepackage{authblk}

\title{A generalized inner product-based wave scattering from an underwater source in a compressible ocean}
\author{ R. Pethiyagoda$^1$, $\,\,$   S. Das$^{2,3}$\thanks{Corresponding author: santudas20072@gmail.com, d.santu@iasst.gov.in}, $\,\,$ B. Wilks$^{1,4}$, $\,\,$ M. H. Meylan$^1$} 

\date{%
	$^1$School of Information and Physical Sciences, University of Newcastle, NSW 2308, Australia\\%
	$^2$Mathematical and Computational Sciences (Physical Sciences Division), Institute of Advanced Study in Science and Technology, Guwahati 781035, India \\%
	$^3$Academy of Scientific and Innovative Research (AcSIR), Ghaziabad - 201002, India\\%
	$^4$School of Mathematical Sciences, Adelaide University, Mawson Lakes SA 5095, Australia\\[2ex]%
	\today
}

\begin{document}
	\maketitle
	
	\begin{abstract}
		Motivated by applications to underwater explosions and volcanic eruptions, this paper considers the evolution of an initial pressure disturbance in the ocean, including effects due to the dynamic and static compression of water and the free surface. In order to solve the equations of motion of a linear compressible ocean, a special inner product is introduced, which allows us to apply self-adjoint operator theory. What results is a Hilbert space in which the acoustic-gravity modes are orthogonal in the generalised sense. This allows the time-domain evolution of the free surface and subsurface pressure field resulting from an initial disturbance to be calculated. Our simulations show initial radial propagation of the pressure pulse and subsequent reflection from the water surface and the rigid ocean floor, eventually leading to horizontal propagation away from the source point. The solutions with and without the inclusion of the static compression are compared, and the effect of static compression is shown to be small but not negligible.  
	\end{abstract}
	
	\section{Introduction}\label{Section1}
	
	The evolution of the pressure field resulting from an initial disturbance in the ocean has been of significant interest to the scientific community for a long time, primarily due to the potential impact of underwater nuclear bomb testing, underwater volcanic eruptions, seismic events, and the associated tsunami hazards. Due to these, there is widespread monitoring of ocean pressure; for example, the Comprehensive Nuclear-Test-Ban-Treaty Organization monitors pressure signals from hydrophone stations to detect possible nuclear tests, and similar devices are also used to detect tsunami waves. Accurate modelling of the resulting pressure signatures allows information about the initial event to be determined, for example, an attempt has been made to estimate the crash site location of Malaysian Airlines Flight 370 by solving an inverse problem \citep{kadri2024underwater}. The accuracy of such inverse problems depends on the accuracy of the forward solution and analysis of the associated wave evolution problem. Obviously, the short--time behaviour of such events may be highly nonlinear and involve the formation of a cavity and subsequent expansion at the initial state. However, after some time (and at some distance), the pressure in the water column can be accurately modelled with a linearised water wave theory that accounts for the effect of dynamic compression (we use this term to distinguish the effect of the local compressibility from the static compression which is the accumulation of compression due to the weight of the overlying water column). 
	
	The inclusion of the compressibility of water is an essential part of our work. Although the compressibility of water is small, its inclusion changes the mathematical structure of the problem so that acoustic-gravity waves (AGW) arise. These are hydroacoustic waves influenced by gravity at low frequency, which are analogous to propagating modes in acoustic or electromagnetic waveguides. These modes capture the effect of the free surface boundary condition and play a pivotal role in the accurate calculation of the pressure of the water column \citep{nosov2018relationship}. It is worth mentioning that in the limit of infinite sound speed in water, the gravity mode transforms into the propagating wave mode under realistic parameter choices, while the AGW modes become evanescent. Following the devastating aftermath of the $2004$ Indian Ocean tsunami, a renewed interest in the application of AGW in tsunami detection mechanisms (through modelling of surface wave propagation) has grown through the works of \cite{nosov1999tsunami,nosov2001nonlinear,nosov2007elastic}, which are based on the well-established fundamental mathematical theory of compressible ocean developed by \cite{longuet1950theory,sells1965effect,miyoshi1954generation,yamamoto1982gravity}. While incompressible water wave theory is most commonly used in practice, the importance of compressibility has been demonstrated by enhanced accuracy in estimating the observed tsunami wave profile \citep{nosov2018relationship}. Following this development, \cite{stiassnie2010tsunamis} and \cite{hendin2013tsunami} computed the far-field profiles of the gravity and AGW modes. Subsequently, there has been a surge in studies investigating different properties of these AGW modes, for example: the production of AGW by interacting gravity waves \citep {kadri2013generation}; triad resonances \citep{kadri2016triad, kadri2016resonant} --- an interaction between two waves resulting in a third wave satifying resonance conditions; their impact on different oceanic processes (e.g., deep-water transport \citep{kadri2014deep}, ice shelf breaking \citep{kadri2012acoustic}, wave dissipation due to sea ice \citep{chen2019dispersion}); and their response to different seafloor characteristics \citep{chierici2010modeling,abdolali2015depth}. Different mathematical techniques are also deployed to solve the associated initial boundary value problem (IBVP) (e.g., rigid but variable bathymetry using depth-integrated mild-slope equations \citep{sammarco2013depth}, Fourier-series expansion \citep{Das2023time-domain}, and eigenfunction expansion \citep{Das2023PoF}), which are essential to address the static compression.
	
	The static compression of the ocean is due to the effect of the slight compressibility of water, causing an increase in the density of the deep ocean. It is only significant for waves which penetrate the deep ocean, and even then, its effect is small. The initial account of static compression was given by \citet{longuet1950theory}. However, until the work of \citet{abdolali2017role} showing the increase of tsunami phase speed at high ocean depth due to static compression, this quantity was rarely included in models due to its presumed insignificant effect on wave propagation. A few contemporary works retained static compression in the mathematical treatment of physical problems to study the behaviour of the gravity mode \citep{kadri2015wave,abdolali2017role, abdolali2019effect,kadri2016triad,kadri2013generation}, most notably by \citet{omira2022global} to match the Hunga Tonga–Hunga Ha`apai volcanic eruption-induced tsunami data using the resonance between the acoustic waves and the surface gravity wave. Variations in water salinity will also affect the density of the ocean over its depth. As such, the vertical profile of sound speed can be complicated. While the common approach is to approximate the sound speed profile as constant,  \cite{michele2020effects} used a perturbation technique on the variable-depth profile of sound speed to derive an analytical expression correct up to third order that governs the evolution of the hydroacoustic waves. More recently, \cite{andrade2025acoustic} used a dynamic approach to study triad resonances and energy exchange between different modes.  Recent developments include a novel Lagrangian method to approach density stratification by linearising the compressible Euler equation \citep{dubois2023acoustic}, and they demonstrated that other models in the literature can be derived from their general proposed model. In the context of tsunami wave generation, \cite{Das2023PoF} solved for the time-dependent evolution due to impulsive forcing from the sea floor, including the effect of static compression. The eigenfunction matching required a special inner product, which is generalised in the present work. This method was later extended to include more general geometries (e.g., a step \citep{Das2024}, a trench \citep{Pethiyagoda_Das_Meylan_2024}, a generic ocean floor \citep{pethiyagoda_arbitrary25}), three-dimensional solutions \citep{das4636389effect}, and surface disturbance-induced wave propagation models \citep{pethiyagoda4677642atmospheric, pethiyagoda_3d_atmos}. These studies augment the works where static compression is dropped \citep{renzi2014hydro} and ocean compressibility is ignored \citep{liu2022water}. We note here that the current work could be extended to include other effects that are non-standard beyond the static compression considered here. Two examples are the effect of the seabed elasticity \citep{eyov2013progressive,abdolali2019effect,kadri2019effect,williams2023acoustic} and more complex static compression using the vertical profile of the sound speed \citep{michele2020effects}. We note that, for the range of parameters we investigate here, the effect of the static compression is small, even for the simplest formulation which we use. 
	
	Many energy-conserving linear wave problems can be recast as an abstract wave equation 
	$\partial_t^2 u+\mathcal{A}u=0$, where $\mathcal{A}$ is a positive self-adjoint operator in a Hilbert space $(\mathcal{H},\langle\cdot,\cdot\rangle)$. In one of the simplest cases, choosing $\mathcal{A}=-c^2\partial_x^2$ with inner product $L_2(\mathbb{R})$ gives the classical wave equation with wave speed $c$. More complicated forms of $\mathcal{A}$ appear in various scenarios such as hydroelasticity \citep{meylan_2002}. 
	While the use of the spectral theorem to obtain the solution as an expansion over the eigenfunctions of $\mathcal{A}$ was long known for the case when $\mathcal{A}$ has a discrete spectrum (Fourier/Sturm Liouville theory) the case when $\mathcal{A}$ has a continuous spectrum began being worked out for the Schr\"odinger equation with the advent of quantum mechanics \citep{povzner1953expansion,ikebe1960eigenfunction}. 
	The theory, which has been extended to the classical wave equation \citep{wilcox1975scattering} and to hydrodynamics for the case of incompressible water wave scattering \citep{hazard2007generalized,hazard2008spectral,meylan_2009}, is known as a generalised eigenfunction expansion.  The theoretical foundation is given by a Gelfand triple, also known as a rigged Hilbert space, which is a mathematical structure used to extend the usual Hilbert space framework to include more generalised functions (i.e., distributions). This allows the treatment of generalised eigenfunctions, i.e., functions $\hat{u}$ satisfying $\mathcal{A}\hat{u}=\omega^2\hat{u}$ but which are not square-integrable (i.e., not belonging to the Hilbert space $\mathcal{H}$). These generalised eigenfunctions are still useful and meaningful in a generalised sense. In general, this theory has been developed in isolation from computation, despite the fact that the simplest example of this structure is the Fourier transform, and its computation by the discrete Fourier transform (often implemented using the fast Fourier transform) is one of the most ubiquitous algorithms. 
	
	It turns out that the theory of generalised eigenfunctions and spectral solutions for continuous spectra can be implemented numerically by applying quadrature to the integrals over the continuous spectrum. This idea was in the work of  \cite{hazard2008spectral,meylan_2009,meylan2009time2, peter2010general}, but appears explicitly in \cite{wilks2024,WILKS2025103421}. As far as we are aware, the present work is the first application of the theory to compressible water wave scattering. A key advantage of the numerical spectral technique is that the generalised eigenfunctions, which satisfy $\mathcal{A}\hat{u}=\omega^2\hat{u}$, are precisely the time harmonic solutions for the angular frequency $\omega$. These single-frequency scattering solutions can be found numerically for a wide range of problems, e.g., variable bathymetry, which would be a relatively straightforward extension of the present work. We note that the ease of determining the spectrum stands in contrast with a more numerical method (e.g., Lanczos Algorithm), in which the operator $\mathcal{A}$ is approximated by a matrix and the spectrum of this matrix is sought. In this case, very many numerical problems arise, especially because the eigenvalues of the underlying operator are continuous. Further problems can arise from the fact that it may be difficult to preserve the self-adjoint property of the matrix if the inner product $\langle\cdot,\cdot\rangle$ is not straightforward to compute (as in the case with the current problem). Finally, we note that the use of these spectral methods is especially useful in problems where there is an initial condition rather than forcing at the boundary.
	
	In this work, we consider the linear initial value problem for an underwater pressure source, including the effect of the free surface. The solution relies on the use of a special Hilbert space. We also extend this formulation to the more challenging case where we include the effect of static compression. 
	The properties of the operator involved in this class of problems are not well studied in the literature. We provide a mathematical structure for the operators involved in a rigid and flat ocean floor via a special inner product. Additionally, the mathematical complexity entangled with static compression makes the problem a difficult one to solve in finite water depth. We aim to show the ease with which the present method can tackle such a complex problem, relaxing any assumptions on the relative water depth, thus making the approach a generalised one. The manuscript is structured as follows. Section 2 presents the two-dimensional mathematical formulation of the physical problem where static compression is ignored, followed by the solution method that involves deriving a generic two-dimensional inner product in \S 3. The time domain simulations are detailed in \S 4. The problem is much simpler if static compression is ignored, and for many scenarios, only this solution will be required, and this is the motivation for giving this in detail. In \S 5, we extend the formulation to static compression and calculate the time-domain simulations of the flow field in this case. We note that the inclusion of static compression complicates the derivation, but the numerical solution takes a similar computational time to the solution without compression. The manuscript concludes with a brief summary in \S 6.
	
	\section{Mathematical formulation without static compression}\label{sec:formulation}
	
	We construct the physical problem in a homogeneous ocean water environment where the standard density variance due to the water column's own weight is ignored, leading to ocean compressibility without static compression (only the dynamic compression is included). The mathematical problem is formulated in a two-dimensional coordinate system $(x,z)$ whose $x$-direction is horizontal and $z$-direction vertically upwards. Linearised water wave theory is used to model the wave scattering induced by an initial pressure in a compressible ocean. The water is considered inviscid and the motion is irrotational, leading to the existence of a time-dependent velocity potential $\Phi(x,z,t)$ throughout the fluid domain such that $\nabla\Phi(x,z,t)=(u(x,z,t),\,v(x,z,t)) =\vec{U}(x,z,t)$ where $u(x,z,t)$ and $v(x,z,t)$ are the fluid velocities in the $x$ and $z$ directions, respectively. The equilibrium free surface and the rigid ocean floor are defined by $z=0$ and $z=-h$, respectively.
	
	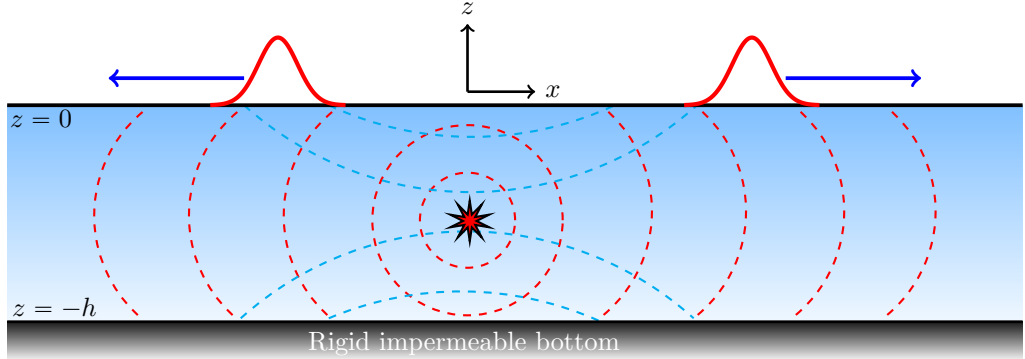
\begin{figure}
		\centering
		\begin{tikzpicture}[scale=0.9]
			\fill [shade,top color=blue!50!cyan!50] (-8.5,-1.4) rectangle (6.5,-5.1); 
			\fill [shade,top color=black!10!black!90] (-8.5,-5.2) rectangle (6.5,-4.6);
			\draw[thick,dashed,red] (-1.7,-3.1) circle (20pt);
			\draw[thick,dashed, red] (-1.7,-3.1) circle (40pt);
			\draw[thick,red,dashed] (0.31,-4.5) arc (-50:50:2cm);
			\draw[thick,red,dashed] (1.7,-4.5) arc (-50:50:2cm);
			\draw[thick,red,dashed] (3.15,-4.5) arc (-50:50:2cm);
			\draw[thick,red,dashed] (4.5,-4.5) arc (-50:50:2cm);
			\draw[thick,red,dashed] (-3.7,-1.5) arc (130:230:2cm);
			\draw[thick,red,dashed] (-5.1,-1.5) arc (130:230:2cm);
			\draw[thick,red,dashed] (-6.5,-1.5) arc (130:230:2cm);
			\draw[thick,cyan,dashed] (-3.75,-1.4) arc (-115:-65:5cm);
			\draw[thick,cyan,dashed] (-3.75,-4.55) arc (113:66:5cm);
			\draw[thick,cyan,dashed] (-5,-1.4) arc (-132:-48:5cm);
			\draw[thick,cyan,dashed] (-5.05,-4.55) arc (132:48:5cm);
			\draw (-8, -1.6) node{$z=0$};
			\draw (-7.8, -4.4) node{$z=-h$};
			\draw [line width=1pt] (-8.5,-4.6)  -- node[below]{} ++(4.5,0) -- node[below] {{\color{white}Rigid impermeable bottom}} ++(4.5,0) -- (6.5,-4.6);
			\draw[line width=0.5mm,color=black] (-8.5,-1.4) -- node[below] {} ++(5.5,0) -- node[below] {} ++(4,0) -- (6.5,-1.4);
			\draw[ultra thick, red] (0,0) plot[domain=-1:1, samples=100]  (\x-4.5,{1*exp(-6*\x*\x )-1.4});
			\draw[ultra thick, red] (0,0) plot[domain=-1:1, samples=100]  (\x+2.5,{1*exp(-6*\x*\x )-1.4});
			\draw [line width=1pt,->] (-1.7,-1.2)  -- (-0.7,-1.2) node(xline)[right] {$x$};
			\draw [line width=1pt,->] (-1.7,-1.2) -- (-1.7,-0.2) node(yline)[above] {$z$};
			\draw[line width=0.5mm,color=blue,->] (3,-1)-- (5,-1);
			\draw[line width=0.5mm,color=blue,->] (-5,-1)-- (-7,-1);
			\tstar{0.1}{0.3}{10}{0}{thick,fill=red, transform canvas={xshift=-1.5cm, yshift=-2.8cm}};
		\end{tikzpicture}
		\caption{Schematic diagram of the physical problem showing wave propagation in a compressible ocean (without including static compression) due to an underwater explosion (red star). Pressure waves generated by the explosion (red dashed lines) are reflected off the sea bed and free surface (blue dashed lines) and induce propagating waves on the fluid surface (solid red lines)}\label{fig:schematic}
	\end{figure} 
	In the absence of extraneous force, the equations of motion can be written as
	\begin{subequations}
		\allowdisplaybreaks  
		\begin{align}
			\vec{U}_t+(\vec{U}\cdot\nabla)\vec{U}&=-\frac{1}{\rho}\nabla p=-\frac{1}{\rho} \frac{\partial p}{\partial\rho}\nabla\rho,\label{0a}\\
			\rho_t+\nabla\cdot(\rho\vec{U})&=0\label{0b},
		\end{align}
	\end{subequations}
	where $\rho$ and $p$ represent the density and pressure, respectively. If $s$ represents the relative change in the density from the undisturbed state (termed as `condensation'), one can write $\rho=\rho_0(1+s)$, where $\rho_0$ represents the water density in the undisturbed state. Additionally, using the definition of the velocity potential and neglecting the higher-order terms, the above equations turn into
	\begin{subequations}\label{pot}
		\allowdisplaybreaks 
		\begin{align}
			\nabla\phi_t&=-\frac{1}{\rho_0} \frac{dp}{d\rho}\nabla s,\\
			s_t&=-\nabla^2\phi.
		\end{align} 
	\end{subequations}
	Denoting $c$ and $K_0$ as sound speed and bulk modulus of the fluid, respectively, and following \cite{lamb1924hydrodynamics}, we get
	\begin{equation}\label{rel}
		\left.\frac{dp}{d\rho}\right|_{\rho=\rho_0} = \frac{K_0}{\rho_0}=c^2.
	\end{equation}
	Eliminating $s$ from \eqref{pot} and making use of \eqref{rel}, we obtain the governing equation as
	\begin{equation*}
		\Phi_{tt}=c^2\nabla^2\Phi.
	\end{equation*}
	Hence, the IBVP, as given by \citep{yamamoto1982gravity, stiassnie2010tsunamis}, is
	\begin{subequations}\label{BVP_time_in}
		\allowdisplaybreaks
		\begin{align}
			\Phi_{tt}&=c^2\nabla^2\Phi,\quad &(x,z)\in\Omega,\label{a}\\
			\Phi_{tt}+g\Phi_{z}&=0, \quad &z=0,\\
			\Phi_{z}&=0, &z=-h,\label{seabed}\\
			\Phi_x&\to 0, & |x|\to \infty, \label{radiation}
		\end{align}
		subject to the initial conditions
		\begin{align}
			\Phi(x,z,0) &= \Phi_0(x,z),\label{init_1}\\
			\Phi_t(x,z,0)&=-\frac{P_0(x,z)}{\rho},\label{init_2}
		\end{align}   
	\end{subequations}
	where $c$ represents the speed of sound in sea water, $g$ is acceleration due to gravity, $\Omega=\{(x,z): |x|<\infty,-h<z<0\}$ is the fluid domain, and $\Phi_0$ and $P_0$ are the initial potential and hydrodynamic pressure of the fluid, respectively. 
	
	We consider this problem in the context of self-adjoint operator theory. We rewrite (\ref{a}) as
	\begin{equation*}
		\Phi_{tt} - c^2\nabla^2\Phi=0,
	\end{equation*}
	and find an inner product with respect to which the operator $-c^2\nabla^2$ is self-adjoint subject to the boundary conditions. To find this inner product, we use Green's second identity in the following form:
	\begin{subequations}
		\begin{align}
			\iint_\Omega \left(\nabla^2\Phi\,\Psi-\Phi\,\nabla^2\Psi\right){\rm d}V&=\oint_{\partial\Omega}\left(\Phi_n\,\Psi-\Phi\,\Psi_n\right){\rm d}S,
		\end{align}
		where $\Phi$ and $\Psi$ are the two solutions of the IBVP \eqref{a}-\eqref{radiation}. After rearranging the terms, we get
		\allowdisplaybreaks
		\begin{align}
			\iint_\Omega \nabla^2\Phi\,\Psi{\rm d}V-\oint_{\partial\Omega}\Phi_n\,\Psi{\rm d}S=\iint_\Omega\Phi\,\nabla^2\Psi{\rm d}V-\oint_{\partial\Omega}\Phi\,\Psi_n{\rm d}S.
		\end{align}
		Taking $\Omega$ to be the region $(-L,L)\times (-h,0)$ gives, in the limit as $L\to\infty$,
		\begin{align}
			&\iint_\Omega \nabla^2\Phi\,\Psi{\rm d}V-\int_{-\infty}^\infty \Phi_z\,\Psi\,{\rm d}x -\lim_{x\rightarrow\infty}\int_{-h}^0\Phi_x\,\Psi\,{\rm d}z+\lim_{x\rightarrow-\infty}\int_{-h}^0\Phi_x\,\Psi\,{\rm d}z \nonumber\\
			&=\iint_\Omega \nabla^2\Psi\,\Phi{\rm d}V-\int_{-\infty}^\infty \Psi_z\,\Phi\,{\rm d}x -\lim_{x\rightarrow\infty}\int_{-h}^0\Psi_x\,\Phi\,{\rm d}z+\lim_{x\rightarrow-\infty}\int_{-h}^0\Psi_x\,\Phi\,{\rm d}z,
		\end{align}
		where we have used \eqref{seabed}. Finally, use of \eqref{radiation} gives
		\begin{align}
			\iint_\Omega \nabla^2\Phi\,\Psi{\rm d}V+\frac{c^2}{g}\int_{-\infty}^\infty \nabla^2\Phi\,\Psi\,{\rm d}x=\iint_\Omega \Phi\,\nabla^2\Psi{\rm d}V+\frac{c^2}{g}\int_{-\infty}^\infty \Phi\,\nabla^2\Psi\,{\rm d}x.\label{eq:Greens_identity_final}
		\end{align}
	\end{subequations}
	Here we used the condition  on $\Phi$ and $\Psi$ and a combination of (\ref{BVP_time_in}a) and (\ref{BVP_time_in}b) that leads to $\Phi_z=-\frac{c^2}{g}\nabla^2\Phi$.
	
	We define the Hilbert space to be the space equipped with the inner product
	\begin{align}\label{inner_product}
		\langle\Phi,\Psi\rangle_{\mathcal{E}}=\int_\Omega \Phi(x,z)\overline{\Psi(x,z)}\,{\rm d}x{\rm d}z+\frac{c^2}{g}\int_{-\infty}^\infty\Phi(x,0)\overline{\Psi(x,0)}\,{\rm d}x.
	\end{align}
	for $\Phi$, $\Psi\in\mathcal{E}$. Self-adjoint operator theory requires an inner product space in which the operator has a symmetry, as we have just shown, and also that the domain of the operator and its adjoint are the same. The latter point is rather subtle, but in our case, this will follow when the function (or solution) space is closed. No formal proof of this is given.  It is worth noting that the operator and the Hilbert space are closely related and that when we close the function space, this closure occurs in the norm defined by our inner product. This function space is also the closure of the eigenfunctions of our operator, which we use to make our numerical calculations. In what follows, we use the inner product to calculate all projections, which is critical to the solution we develop. 
	
	\section{Solution technique}\label{sec:solution_techniques}
	\noindent Since the operator $-c^2\nabla^2$ is assumed to be self-adjoint, we know that its spectrum is real. In fact, it has a continuous spectrum occupying the entire positive real line. To find the generalised eigenfunctions, we solve 
	\begin{subequations}
		\begin{align}\label{helmholtz}
			c^2\nabla^2\phi &= -\omega^2\phi,&(x,z)\in\Omega,\\
			-\omega^2\phi+g\phi_z&=0,& z=0,\\
			\phi_n&=0,& z=-h,
		\end{align}
	\end{subequations}
	subject to an incident wave condition. Solutions of \eqref{helmholtz} are nothing more than the frequency domain solutions found by assuming time harmonicity, namely
	\begin{equation}
		\Phi(x,z,t) = \Re\{\phi(x,z) \mathrm{e}^{-{\rm i}\omega t}\},
	\end{equation}
	where $\omega$ is the angular frequency. 
	
	Using separation of variables, we obtain explicit expressions for the solutions of \eqref{helmholtz} of the form
	\begin{align} 
		\phi^{\pm}_n(x,z;k)=\mathrm{e}^{{\pm\rm i}k x}f_n(z;k),
	\end{align}
	for $n\in\mathbb{N}$ and $k\in\mathbb{R}$, where the depth dependent functions satisfy the eigenvalue problem
	\begin{subequations}\label{f_eq}
		\begin{equation}f_n''(z;k)=\mu_n^2f_n(z;k),
		\end{equation}
		subject to the boundary conditions
		\begin{align}
			gf_n'(0;k)&=c^2(k^2-\mu_n^2)f_n(0;k),\\
			f_n'(-h;k)&=0,
		\end{align}
	\end{subequations}
	where the prime denotes differentiation with respect to $z$. Equation \eqref{f_eq} has a discrete spectrum of eigenvalues $\mu_n$, indexed by $n\in\mathbb{N}_0$,   satisfying the following dispersion relation
	\begin{align}
		\mu_n(k)\tanh{\mu_n(k) h}+\frac{c^2}{g^2}\mu_n(k)^2=\frac{c^2k^2}{g^2},\label{eq:dispersion_relation}
	\end{align}
	with $\mu_0(k)$ being the positive real root and $\mu_n(k)$ being of the form $\rm i\tilde{\mu}_n(k)$ for $n\geq 1$, where $\tilde{\mu}_n(k)$ are real, positive and in ascending order with respect to the index $n$. In addition, $k_n$ are real positive for all progressive acoustic-gravity wave modes and imaginary for evanescent modes. Note that while $\mu_n(k)$ depends on both $n$ and $k$, we will omit the argument $k$ when this is clear from context. The associated vertical eigenfunctions are written as
	\begin{align}
		f_n(z;k)=&\frac{\displaystyle \cosh{\mu_n(z+h)}}{\cosh{(\mu_n h)}}.
	\end{align}
	With reference to \eqref{helmholtz}, the time harmonic angular frequency associated with the mode $\phi_n(\cdot,\cdot;k)$ is
	\begin{equation}
		\omega_n(k)^2=  c^2(k^2-\mu_n^2).\label{eq:dispersion_relation2}
	\end{equation}
	
	Using the spectral theorem, the general solution of the time domain problem \eqref{BVP_time_in} can be written as the following series of integrals over the continuous wavenumber
	\begin{equation}
		\Phi(x,z,t)=  \sum_{n=0}^{\infty}\int_{0}^\infty 
		\left(C_n^-(t,k)\phi_n^{-}(x,z;k) {\rm d}k
		+  C_n^+(t,k)
		\phi_n^{+}(x,z;k)\right){\rm d}k.
	\end{equation}
	Substitution of the above into \eqref{a} then gives
	\begin{multline}
		\sum_{n=0}^{\infty}\int_{0}^\infty -\omega_n(k)^2
		\left(C_n^-(t,k)\phi_n^{-}(x,z;k) {\rm d}k
		+  C_n^+(t,k)
		\phi_n^{+}(x,z;k)\right){\rm d}k\\=\sum_{n=0}^{\infty}\int_{0}^\infty 
		\left(C_n^{-\prime\prime}(t,k)\phi_n^{-}(x,z;k) {\rm d}k
		+  C_n^{+\prime\prime}(t,k)
		\phi_n^{+}(x,z;k)\right){\rm d}k,
	\end{multline}
	where the prime denotes differentiation with respect to $t$. Using the following orthogonality results (which we prove in Appendix \ref{sec:orthogonality}):
	\begin{subequations}\label{orthonormality_conds}
		\begin{align}
			\langle\phi_n^\pm(\cdot,\cdot,k),\phi_p^\pm(\cdot,\cdot,\tilde{k})\rangle &= D_p(k)\delta(k-\tilde{k})\delta_{np}&k,\tilde{k}>0,\quad n,p \in\mathbb{N},\\
			\langle\phi_n^\pm(\cdot,\cdot,k),\phi_p^\mp(\cdot,\cdot,\tilde{k})\rangle &= 0&k,\tilde{k}>0,\quad n,p \in\mathbb{N},
		\end{align}
	\end{subequations}
	noting that (\ref{orthonormality_conds}a) is the inner product between waves travelling in the same direction and (\ref{orthonormality_conds}b) is the inner product between waves travelling in opposite directions. Here, in which $D_p(k)$ are normalisation coefficients, $\delta(\cdot)$ is the Dirac delta and $\delta_{np}$ is the Kronecker delta, we obtain
	\begin{equation}
		C_n^{\pm\prime\prime}(t,k)=-\omega_n(k)^2C_n^\pm(t,k).
	\end{equation}
	Solutions to the above ordinary differential equation can be written as
	\begin{equation}
		C_n^\pm(t,k)=A_n^\pm(k)\cos(\omega_n(k)t)+B_n^\pm(k)\frac{\sin(\omega_n(k)t)}{\omega_n(k)},
	\end{equation}
	whence
	\begin{multline}\label{general_td_sol}
		\Phi(x,z,t)=  \sum_{n=0}^{\infty}\int_{0}^\infty 
		\left(A_n^-(k)\phi_n^{-}(x,z;k)+A_n^+(k)
		\phi_n^{+}(x,z;k)\right) \cos(\omega_n(k) t){\rm d}k \\
		+ \sum_{n=0}^{\infty}\int_{0}^\infty 
		\left(B_n^-(k)\phi_n^{-}(x,z;k)+B_n^+(k)
		\phi_n^{+}(x,z;k)\right) \frac{\sin(\omega_n(k) t)}{\omega_n(k)}{\rm d}k.
	\end{multline}
	To obtain the particular solution to the IBVP \eqref{BVP_time_in}, the coefficient functions $A_n^\pm$ and $B_n^\pm$ must be determined from the initial potential and pressure of the fluid, $\Phi_0$ and $P_0$, respectively. Applying \eqref{orthonormality_conds} to \eqref{general_td_sol} for $t=0$ yields
	\begin{subequations}   
		\begin{align}                 
			\langle\Phi(\cdot,\cdot,0),\phi_p^\pm(\cdot,\cdot, \tilde{k}) \rangle_{\mathcal{E}} &=\sum_{n=0}^{\infty} \int_{0}^\infty A_n^-(k)\langle\phi_n^-(\cdot,\cdot,k), \phi_p^\pm(\cdot,\cdot,\tilde{k})\rangle_{\mathcal{E}}  +A_n^+(k)\langle\phi_n^-(\cdot,\cdot,k),\phi_p^\pm(\cdot,\cdot,\tilde{k})\rangle_{\mathcal{E}} \,{\rm d}k, \nonumber\\
			&=D_p(\tilde{k})A_p^\pm(\tilde{k})
		\end{align}
		Similarly
		\begin{equation} \langle\Phi_t(\cdot,\cdot,0),\phi_p^\pm(\cdot,\cdot,\tilde{k})\rangle_{\mathcal{E}} =D_p(\tilde{k})B_p^\pm(\tilde{k}).
		\end{equation}
	\end{subequations}
	Now we apply the initial conditions \eqref{init_1} and \eqref{init_2} and take inner products with respect to $\phi_p^\pm$ to get 
	\begin{subequations}
		\begin{align}
			\langle \Phi_0(x,z),\phi_p^\pm(\cdot,\cdot,k)\rangle_{\mathcal{E}} &=\langle\Phi(\cdot,\cdot,0),\phi_p^\pm(\cdot,\cdot,k) \rangle_{\mathcal{E}}=D_pA_p^\pm(k),\\
			-\frac{\langle P_0(x,z),\phi_p^\pm(\cdot,\cdot,k)\rangle_{\mathcal{E}} }{\rho}&=\langle\Phi_t(\cdot,\cdot,0),\phi_p^\pm(\cdot,\cdot,k)\rangle_{\mathcal{E}}=D_pB_p^\pm(k).
		\end{align}
	\end{subequations}
	Then the coefficient functions $A_p^\pm(k)$ and $B_p^\pm(k)$ are given by
	\begin{equation}\label{recover_coeffs}
		A_p^\pm(k)=\frac{\langle \Phi_0(x,z),\phi_p^\pm(\cdot,\cdot,k)\rangle_{\mathcal{E}} }{D_p} \quad\mbox{and}\quad B_p^\pm(k)=-\frac{\langle P_0(x,z),\phi_p^\pm(\cdot,\cdot,k)\rangle_{\mathcal{E}} }{\displaystyle\rho D_p}.
	\end{equation}
	Hence, we obtain the potential function as
	\begin{multline}\label{eq:phi_actual}
		\Phi(x,z,t)=\sum_{n=0}^{\infty}\int_{0}^\infty\left(\frac{\langle \Phi_0(x,z),\phi_n^-\rangle_{\mathcal{E}}}{D_p}  \phi_n^-+\frac{\langle \Phi_0(x,z),\phi_n^+\rangle_{\mathcal{E}}}{D_p}  \phi_n^+\right)\cos(\omega_n t)\,{\rm d}k\\
		-\frac{1}{\rho}\sum_{n=0}^{\infty}\int_{0}^\infty\left(\frac{\langle P_0(x,z),\phi_n^-\rangle_{\mathcal{E}}}{D_p}  \phi_n^-+\frac{\langle P_0(x,z),\phi_n^+\rangle_{\mathcal{E}}}{D_p}  \phi_n^+\right)\frac{\sin(\omega_n t)}{\omega_n}\,{\rm d}k.
	\end{multline}
	We note at this point some special features of this solution for $\Phi$. The solution can be calculated at any time from the initial conditions without reference to the solution at any other time. Moreover, the difficulty of solving for any time is equivalent as the method does not rely on temporal iteration. In particular, we compute the solution at $t=0$, which provides a validation of the method through comparison with the initial condition. We note that, provided that the $\phi_n^{\pm}$ are correct, then $\Phi$ must satisfy the governing equations, and if the potential matches at $t=0$, we have calculated the solution correctly. 
	
	\section{Time-domain simulation}\label{sec:2d_time_domain}
	We will now compute time-domain solutions of the internal pressure 
	\begin{equation}
		P(x,z,t)=-\rho\Phi_t(x,z,t),\label{eq:pressure_solution}
	\end{equation}
	and free-surface
	\begin{equation}
		\eta(x,t)=-\frac{1}{g}\Phi_t(x,0,t)=\frac{1}{\rho g}P(x,0,t).\label{eq:eta}
	\end{equation}
	Noting that when computing the pressure over the full $z$-domain, we recover the surface as a byproduct.
	\subsection{Numerical evaluation}
	When evaluating (\ref{eq:pressure_solution}), care must be taken due to the potentially highly oscillatory nature of the integrand (for large $x$ or $t$). We reformulate (\ref{eq:phi_actual}) into the form
	\[
	\Phi(x,z,t) = \sum_{n=0}^\infty I_n^{(1)}-\frac{1}{\rho}\sum_{n=0}^\infty I_n^{(2)},
	\]
	where
	\begin{align*}
		I_n^{(1)}&=\sum_{\pm}\sum_{j=\{0,1\}}(-1)^j\int_{0}^\infty g_{n\pm}^{(1)}e^{{\rm i}(\pm kx+(-1)^j\omega_nt)}\,{\rm d}k ,\\
		I_n^{(2)}&=\sum_{\pm}\sum_{j=\{0,1\}}\int_{0}^\infty g_{n\pm}^{(2)}e^{{\rm i}(\pm kx+(-1)^j\omega_nt)}\,{\rm d}k ,
	\end{align*}
	and   
	\[
	g_{n\pm}^{(1)}=\frac{\omega_n\langle \Phi_0(x,z),\phi_n^\pm\rangle_{\mathcal{E}}f_n(z)}{2D_p}, \quad\quad g_{n\pm}^{(2)}=\frac{\langle P_0(x,z),\phi_n^\pm\rangle_{\mathcal{E}}f_n(z)}{2D_p}.
	\]
	The infinite sum is truncated at $N$ modes while the semi-infinite integral is truncated at $k=k_\mathrm{max}$ and discretised uniformly using spacing $\Delta k$. The choice of $N$, $k_\mathrm{max}$, and $\Delta k$ depends on the initial conditions and will be discussed in \S\ref{sec:explosion_non_static}. In this paper, we take $N=100$ as that was sufficient to reproduce the initial condition. Similarly, the semi-infinite integral is truncated at $k=0.2$ and discretised uniformly using $1001$ points.
	
	This formulation allows us to implement a Filon-type quadrature method \citep{Dahlquist_Björck_2008,filon_quadrature_1930} detailed in Appendix \ref{sec:filon-type}. Further efficiency can be gained by recognising that
	\[g_{+}e^{{\rm i}(kx+\omega_nt)}=\overline{g_{-}e^{-{\rm i}(kx+\omega_nt)}},\]
	halving the number of terms in the summations that must be computed.
	
	\subsection{Underwater pressure disturbance}\label{sec:explosion_non_static}
	As an example, we will consider an underwater pressure disturbance as the initial condition (generated by an underwater explosion, for example). The fluid is assumed initially at rest, $\Phi_0(x,z)=0$, with the initial pressure distribution
	\begin{equation}\label{eq:inital_pressure}
		P_0(x,z) = \hat{P}e^{\displaystyle-\pi^2\frac{(x-x_c)^2+(z-z_c)^2}{\sigma^2}},
	\end{equation}
	where $\hat{P}$ is the maximum pressure, $(x_c,z_c)$ is the centre of the pressure distribution and $\sigma$ is the characteristic width of the pressure. For the following results, we choose $\hat{P}=10^6$ Pa, $(x_c,z_c)=(0,-2000)$ m, $g=9.81$m/s$^2$, $c=1450$m/s, $h=4000$m and $\sigma=200$m.
	
	The numerical parameters $N$, $k_\mathrm{max}$, and $\Delta k$ are chosen by ensuring the initial condition can be reproduced with sufficient accuracy. We observe the maximum difference between the numerical solution and true initial condition over different parameter choices (see Appendix \ref{app:error} for details). Based on this parameter sweep, we choose $N=100$, $k_\mathrm{max}=0.2$, and $\Delta k=0.0002$.
	
	\begin{figure}
		\centering    
		(a)\hfill
		\begin{subfigure}{}
			\includegraphics[width=0.98\textwidth]{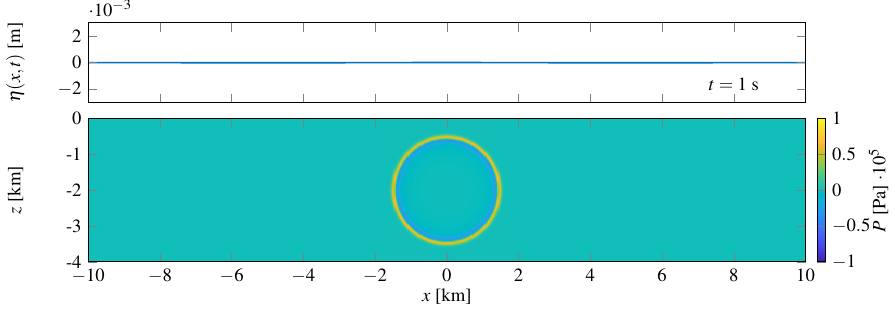}
		\end{subfigure}\\
		(b)\hfill
		\begin{subfigure}{}
			\includegraphics[width=0.98\textwidth]{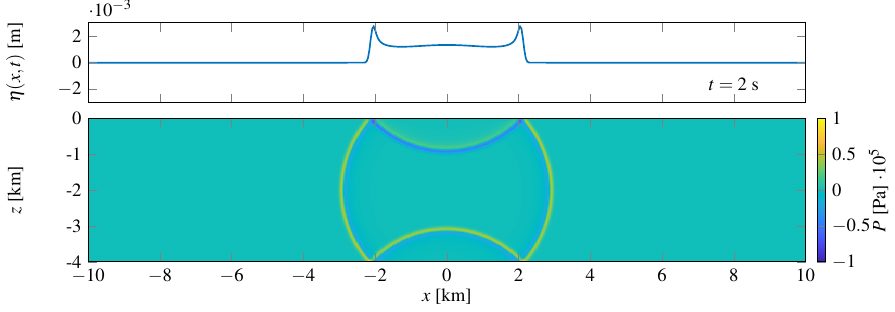}
		\end{subfigure}\\
		(c)\hfill
		\begin{subfigure}{}
			\includegraphics[width=0.98\textwidth]{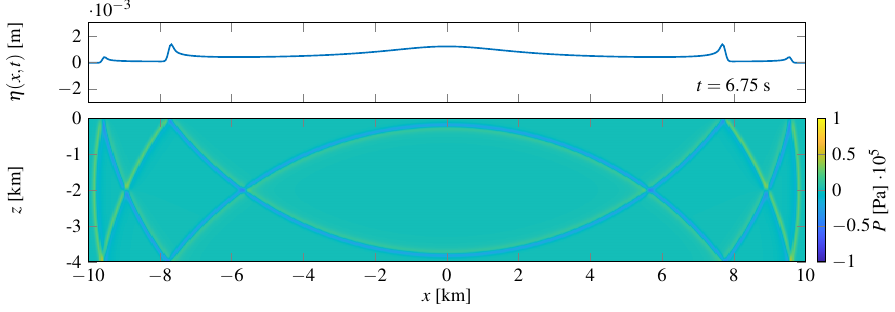}
		\end{subfigure}\\
		\caption{Compression waves caused by an initial Gaussian pressure distribution centred at $(0,-2000)$.
		}
		\label{fig:underwater_gaussian}
	\end{figure} 
	Figure \ref{fig:underwater_gaussian} shows the solutions at $t=1$, $2$, and $6.75$s. We can see that the initial pressure spreads radially away from the initial location at the speed of sound. The expanding wave reflects off both the bottom and the free surface, the phase of the pressure wave reversing as it reflects off the surface (Figure \ref{fig:underwater_gaussian}b). This process continues as time evolves, resulting in multiple pressure waves propagating away from the pressure origin. The effect this has on the free surface is the creation of wave crests above the initial pressure location that propagate away. The free-surface elevation between the crests will also be greater than at equilibrium. Full animation is given as Movie $1$.

	\section{Inclusion of static compression}\label{static_compression_sec}
	
	In this section, we include (mild) density variance of the ocean water, the effect which is known in the literature as static compression, into the mathematical formulation, and develop the solution using a similar approach to that used in the previous section. The rigorous derivation of the governing equations can be found in the works of \cite{longuet1950theory}, \cite{miyoshi1954generation}, and \cite{sells1965effect}. Note that in the subsequent mathematical treatment, the density $\rho$ represents its value at the surface, which is treated as the baseline density. The hydrostatic density is $\rho e^{-\gamma z}$
	\citep{abdolali2017role}, where the effect of the variance in the density away from the surface is reflected through the term $\gamma$, which depends on the speed of sound in the water and is defined by $\gamma = g/c^2$. This inclusion inherently modifies the underlying governing equation which we briefly show now. 
	
	Assuming $\rho_0(z)$ and $p_0(z)$ being  the undisturbed water density and pressure, respectively, the fundamental equations of motion up to the first order yield \citep{bondi1947waves}
	\begin{subequations}
		\begin{align}
			&\phi_t=-\int \frac{dp}{\rho}+gz,\label{bern}\\
			&\rho_t+\nabla\cdot(\rho_0\nabla\phi)=0. \label{dens}
		\end{align}    
	\end{subequations}
	According to \cite{longuet1950theory}, pressure and density are related by the relationship
	\begin{equation}
		\frac{\partial p}{\partial \rho}= c^2,
	\end{equation}
	which modifies \eqref{bern} to 
	\begin{equation}
		\phi_t=-c^2\log\rho +gz.\label{bern2}
	\end{equation} 
	For the undisturbed state, one can obtain from \eqref{bern2}
	\begin{equation}
		\rho_0= \hat{\rho}_0e^{gz/c^2} =\hat{\rho}_0e^{\gamma z},
	\end{equation}
	where $\hat{\rho}_0$ represents undisturbed density at $z=0$. Using $\rho=\rho_0(1+s)$, \eqref{bern2} and \eqref{dens} become
	\begin{subequations}
		\begin{align}
			&\frac{\partial\phi}{\partial t}=-c^2s,\\
			& \frac{\partial s}{\partial t}+\nabla^2\phi + \gamma\phi_z=0.
		\end{align}    
	\end{subequations}
	Eliminating $s$ yields the governing equation
	\begin{equation}
		\nabla^2\Phi= \frac{1}{c^2}\Phi_{tt}+\gamma\Phi_{z}.
	\end{equation}
	The effect of static compression is small for realistic oceans \citep{abdolali2019effect}, as we shall see, and it significantly complicates the formulation. We believe that including it is worthwhile, first because it allows accurate calculation of its effect and second because it highlights how powerful the special inner product-based method is and how it can be extended to highly non-trivial cases. We note that the complexity arises in the theory, but that the computational difficulty is not greatly increased.
	
	\subsection{Mathematical Solution with Static Compression}
	The modified IBVP is given by
	\allowdisplaybreaks
	\begin{subequations}\label{BVP_time_in_stat}
		\begin{align}
			\nabla^2\Phi&= \frac{1}{c^2}\Phi_{tt}+\gamma\Phi_{z},\quad &(x,z)\in\Omega,\label{st}\\
			\Phi_{tt}+g\Phi_{z}&=0, \quad &z=0,\label{fs_st}\\
			\Phi_{z}&=0, &z=-h,\label{seabed_st}\\
			\Phi_{x}&\to 0, &|x|\to\infty,\label{ff_st}
		\end{align}
	\end{subequations}
	subject to the initial conditions
	\allowdisplaybreaks
	\begin{align}
		\Phi(x,z,0) &= 0,\\
		\Phi_t(x,z,0)&=-\frac{P_0(x,z)}{\rho e^{-\gamma z}}.\label{init2}
	\end{align}
	In order to identify an appropriate inner product, we apply Green's second identity to the functions $\hat{\Phi}=e^{-\gamma z/2}\Phi$ and $\hat{\Psi}=e^{-\gamma z/2}\Psi$, where $\Phi$ and $\Psi$ satisfy \eqref{BVP_time_in_stat}, in the form
	\begin{subequations}
		\allowdisplaybreaks
		\begin{align}
			\iint_\Omega \left(\nabla^2\hat{\Phi}\,\hat{\Psi}-\hat{\Phi}\,\nabla^2 \hat{\Psi}\right){\rm d}V=\oint_{\partial\Omega}\left(\hat{\Phi}_n\, \hat{\Psi}-\hat{\Phi}\,\hat{\Psi}_n\right){\rm d}S.\nonumber
		\end{align}
		The choices of $\hat{\Phi}$ and $\hat{\Psi}$ ensure elimination of the first order derivative term in \eqref{st}. Now
		taking $\Omega$ to be the region $(-L,L)\times(-h,0)$ gives, in the limit as $L\to\infty$
		\begin{align}
			\iint_\Omega e^{-\gamma z}\left[\left(\nabla^2\Phi-\gamma\Phi_z\right)\Psi-\Phi\left(\nabla^2\Psi-\gamma\Psi_z\right)\right]{\rm d}V = \int_{-\infty}^\infty \left.\left(\Phi_z\,\Psi-\Phi\,\Psi_z\right)\right|_{z=0}{\rm d}x,\nonumber
		\end{align}
		where we have used \eqref{seabed_st} and \eqref{ff_st}. Subsequently
		\begin{align}
			&\iint_\Omega e^{-\gamma z}\left(\nabla^2\Phi-\gamma\Phi_z\right)\Psi{\rm d}V-\int_{-\infty}^\infty \left.\Phi_z\,\Psi\right|_{z=0}{\rm d}x=\iint_\Omega e^{-\gamma z}\Phi\left(\nabla^2\Psi-\gamma\Psi_z\right){\rm d}V\nonumber\\
			&\hspace{3in}-\int_{-\infty}^\infty \left.\Phi\,\Psi_z\right|_{z=0}{\rm d}x.
		\end{align}
		Note that the combination of \eqref{st} and \eqref{fs_st} yields $\Phi_z=-\frac{c^2}{g}(\nabla^2\Phi-\gamma\Phi_z)$ on $z=0$, which further modifies the above expression into  
		\begin{align}
			\iint_\Omega e^{-\gamma z}\mathcal{A}\Phi\Psi{\rm d}V+\frac{c^2}{g}\int_{-\infty}^\infty \left.\mathcal{A}\Phi\,\Psi\right|_{z=0}{\rm d}x&=\iint_\Omega e^{-\gamma z}\Phi\mathcal{A}\Psi{\rm d}V+\frac{c^2}{g}\int_{-\infty}^\infty \left.\Phi\,\mathcal{A}\Psi\right|_{z=0}{\rm d}x,
		\end{align}
	\end{subequations}
	where $\mathcal{A}=\nabla^2-\gamma\partial_z$. Consequently, this operator is symmetric provided we choose the inner product to be
	\begin{equation}
		\langle\Phi,\Psi\rangle_{\mathcal{E}} =\int_\Omega e^{-\gamma z}\Phi(x,z)\overline{\Psi(x,z)}\,{\rm d}x{\rm d}y+\frac{c^2}{g}\int_{-\infty}^\infty\Phi(x,0)\overline{\Psi(x,0)}\,{\rm d}x.
	\end{equation}
	In a way similar to what is mentioned in \textsection\ref{sec:solution_techniques}, we expand the time domain solution in the generalised eigenfunctions of the underlying differential operator. These eigenfunctions satisfy
	\begin{align}
		\nabla^2\phi&= -\frac{\omega^2}{c^2}\phi+\gamma\phi_{z},\quad &(x,z)\in\Omega,\\
		-\omega^2\phi+g\phi_{z}&=0, \quad &z=0,\\
		\phi_{z}&=0, &z=-h,
	\end{align}
	and are of the form
	\begin{equation}
		\phi_n^\pm(x,z;k)=e^{{\rm i}kx}f_n(z;k).
	\end{equation}
	In the case of static compression, the depth-dependent functions $f_n(z,k)$ are more complicated. They satisfy the eigenvalue problem
	\begin{subequations}\label{depth_dep_eig_problem_static}
		\begin{align}
			f_n''(z;k)-\gamma f_n'(z;k) &= \left(\mu_n^2-\frac{\gamma^2}{4}\right)f_n(z;k),\\
			f_n'(-h;k)&=0,\\
			f'(0;k)&=\frac{c^2}{g}\left(k^2+\frac{\gamma^2}{4}-\mu_n^2\right)f(0;k),
		\end{align}
	\end{subequations}
	where the prime denotes differentiation with respect to $z$. For each $k$, the discrete spectrum of eigenvalues $\mu_n$ of \eqref{depth_dep_eig_problem_static} satisfy the following dispersion relation
	\begin{align}\label{eq:dispRel}
		\frac{c^2}{g}\left(k^2+\frac{\gamma^2}{4}-\mu_n^2\right)\left(\displaystyle 1-\left(\frac{ \gamma}{2\mu_n} \right)\tanh{\mu_n h}\right)=\mu_n\displaystyle\left[1- \left( \frac{\gamma}{2\mu_n} \right)^2 \right]\tanh{\mu_n h}.
	\end{align}
	The solutions to \eqref{eq:dispRel} are taken so that $\mu_0(k)\in\mathbb{R}$ and $\mu_n(k)$ are purely imaginary with positive and increasing imaginary part. The angular frequency associated with the mode $\phi_n^\pm(\cdot,\cdot;k)$ is
	\begin{equation}
		\omega_n(k)=c\sqrt{k^2+\frac{\gamma^2}{4}-\mu_n^2.}
	\end{equation}
	Without loss of generality, we set $f(0)=1$, yielding
	\begin{align}
		f_n(z;k)=&\frac{\displaystyle e^{(\gamma z/2)}\left( \frac{\gamma}{2\mu_n}\sinh{\mu_n(z+h)}-\cosh{\mu_n(z+h)} \right)}{\frac{\gamma}{2\mu_n}\sinh{( \mu_n h)}-\cosh{(\mu_n h)}}.
	\end{align}
	As in \textsection\ref{sec:solution_techniques}, the spectral theorem gives a time domain solution of the form
	\begin{multline}\label{eq:phi_og_stat}
		\Phi(x,z,t)=  \sum_{n=0}^{\infty}\int_{0}^\infty 
		\left(A_n^-(k)\phi_n^{-}(x,z;k)+A_n^+(k)
		\phi_n^{+}(x,z;k)\right) \cos(\omega_n(k) t){\rm d}k \\
		+ \sum_{n=0}^{\infty}\int_{0}^\infty 
		\left(B_n^-(k)\phi_n^{-}(x,z;k)+B_n^+(k)
		\phi_n^{+}(x,z;k)\right) \frac{\sin(\omega_n(k) t)}{\omega_n(k)}{\rm d}k,
	\end{multline}
	where the coefficient functions $A_n^\pm$ and $B_n^\pm$ are to be determined from the initial conditions. With orthonormality conditions again being of the form \eqref{orthonormality_conds}, albeit with different normalisation functions $D_p(k)$, the coefficients are recovered as
	\begin{equation}\label{recover_coeffs_stat}
		A_p^\pm(k)=\frac{\langle \Phi_0(x,z),\phi_p^\pm(\cdot,\cdot,k)\rangle_{\mathcal{E}} }{D_p} \quad\mbox{and}\quad B_p^\pm(k)=-\frac{\langle P_0(x,z)/\rho e^{-\gamma z},\phi_p^\pm(\cdot,\cdot,k)\rangle_{\mathcal{E}} }{\displaystyle D_p},
	\end{equation}
	which is precisely in the form of \eqref{recover_coeffs}. To finalise the mathematical solution in the case of static compression, it remains to compute the normalisation functions $D_p(k)$. This is done in Appendix \ref{normalisation_appendix}. We note that (\ref{eq:phi_og_stat}) and (\ref{recover_coeffs_stat}) reduce to (\ref{eq:phi_actual}) and (\ref{recover_coeffs}), respectively, by setting $\gamma=0$.

	\subsection{Time-domain simulations with static compression}
\begin{figure}[h!]
		\centering    
		(a)\hfill
		\begin{subfigure}{}
			\includegraphics[width=0.98\textwidth]{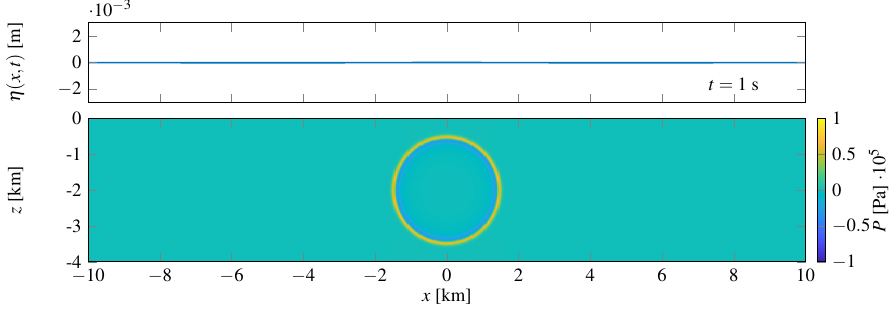}
		\end{subfigure}\\
		(b)\hfill
		\begin{subfigure}{}
			\includegraphics[width=0.98\textwidth]{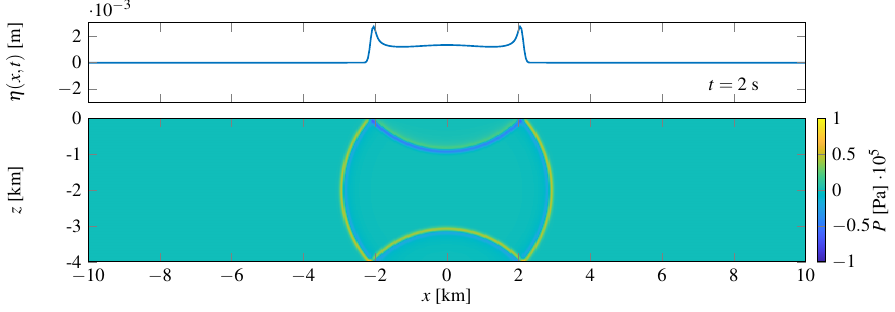}
		\end{subfigure}\\
		(c)\hfill
		\begin{subfigure}{}
			\includegraphics[width=0.98\textwidth]{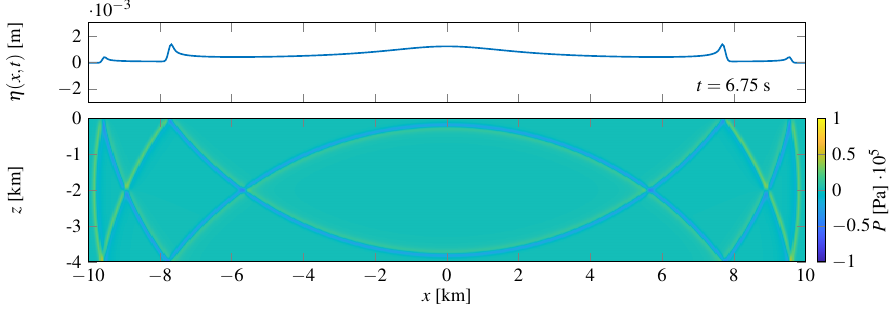}
		\end{subfigure}\\
		\caption{Compression waves caused by an initial Gaussian pressure distribution centred at $(0,-2000)$, including the effects of static compression. In each panel, the line graph represents the free surface displacement, and the colour plot represents the hydrodynamic pressure distribution below the surface.}
		\label{fig:underwater_gaussian_static}
	\end{figure}

\begin{figure}[!h]
		\centering    
		\includegraphics[width=0.98\textwidth]{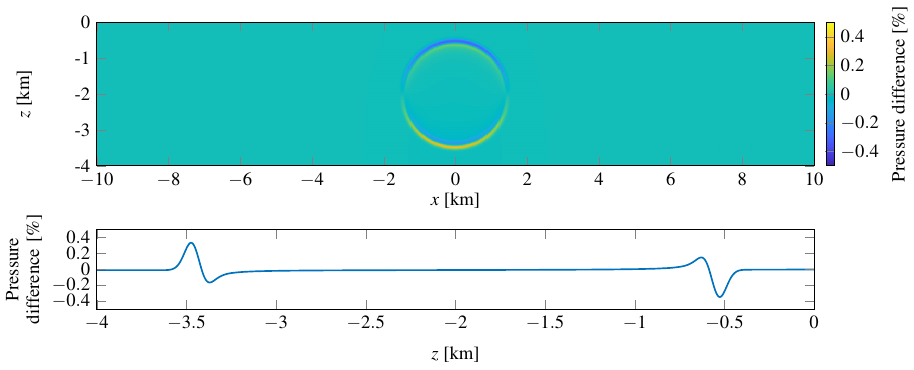} 
		\caption{A plot of the pressure difference between Figure \ref{fig:underwater_gaussian}(a) and Figure \ref{fig:underwater_gaussian_static}(a) as a percentage of that maximum magnitude of pressure at that time ($t=1$), where a positive difference indicates that the pressure solution including static compression is greater than the pressure solution without static compression. The bottom panel provides a cross-section of the pressure difference at $x=0$.}
		\label{fig:underwater_gaussian_static_difference}
	\end{figure}

	Firstly, we consider the underwater explosion example detailed in \S\ref{sec:explosion_non_static}, to compare with the results in the absence of static compression. We can see in Figure \ref{fig:underwater_gaussian_static} (the full animation is in Movie $2$) that the results are visually indistinguishable from the prior results without static compression (Figure \ref{fig:underwater_gaussian}). To better observe the effects of static compression, we plot the percentage difference 
	\begin{equation}
		\frac{P_\mathrm{static}(x,z,t)-P(x,z,t)}{\max\limits_{x,z}(P_\mathrm{static}(x,z,t),P(x,z,t))},\label{eq:static_diff}
	\end{equation}
	for $t=1$ (Figure \ref{fig:underwater_gaussian_static_difference}), where $P_\mathrm{static}$ is the solution including static compression. We observe that the difference wave has a negative leading front in the upper half wave (i.e., upwards travelling waves) and a positive leading front in the lower half (i.e., downward travelling waves). We can see that the inclusion of static compression, on average, causes the magnitude of pressure above (below) the initial explosion location to decrease (increase) compared with the case with no static compression. It aligns with the fact that water is less (more) dense in the upper (lower) water region measured from the point of explosion when static compression is considered.
	Further, we calculated the maximum of the percentage difference (\ref{eq:static_diff}) over space ($|x|<10^5$,$-h<z<0$) and time ($0<t<20$) for different depths, $h$, where the pressure is centred halfway between the bottom and the surface $(x_c,z_c)=(0,-h/2)$ and for a pressure centre fixed at $(0,-250)$. From the results presented in Table \ref{tab:Num_error_depth}, we can see that the pressure difference increases as the depth decreases, and in the case of the pressure at $z_c=-250$m, there is a minimum in pressure difference due to static compression at roughly $h=1500$m before the difference increases as the depth increases. However, the differences still remain roughly within $\pm1\%$ for all tested depths. Finally, we note that the pressure wave speed has not changed with the addition of static compression. The Movie $3$ provides the full animation for Figure \ref{fig:underwater_gaussian_static_difference}.
	\begin{table}
		\centering
		\begin{tabular}{ccc}
			\multirow{3}{*}{Depth ($h$)} & Percentage pressure & Percentage pressure \\
			&difference for&difference for\\
			&$z_c=-h/2$ (\%)&$z_c=-250$ (\%)\\
			500 & 1.2415 & 1.2415\\
			1000 & 0.6565 & 0.6688\\
			1500 & 0.5042 & 0.5877\\
			2000 & 0.4572 & 0.6535\\
			2500 & 0.5306 & 0.6978\\
			3000 & 0.4935 & 0.8619\\
			3500 & 0.5170 & 0.8744\\
			4000 & 0.5988 & 1.0047\\
		\end{tabular}
		\caption{The maximum percentage pressure difference (\ref{eq:static_diff}) for different depths, $h$, where the pressure is centred either halfway between the bottom and the surface $(x_c,z_c)=(0,-h/2)$, or at $(0,-250)$.}
		\label{tab:Num_error_depth}
	\end{table}
	
	We will now consider two different initial pressure distributions that have the same surface displacement but drastically different subsurface profiles. The first initial pressure distribution is the Gaussian distribution (\ref{eq:inital_pressure}) centred at the free-surface $(x_c,z_c)=(0,0)$, and the second is a one-dimensional Gaussian with the same $x$ behaviour as (\ref{eq:inital_pressure}),
	\begin{equation}\label{eq:alt_inital_pressure}
		P_1(x,z) = \hat{P}\exp\left({\displaystyle-\pi^2\frac{(x-x_c)^2}{\sigma^2}}\right).
	\end{equation}
	
\begin{figure}[!h]
		\centering    
		(a)\hfill
		\begin{subfigure}{}
			\includegraphics[width=0.98\textwidth]{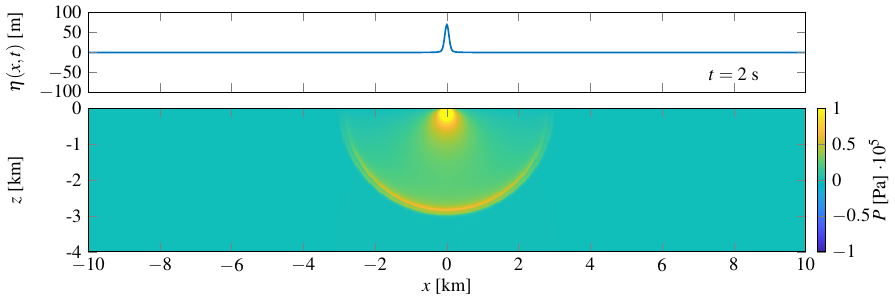}
		\end{subfigure}\\
		(b)\hfill
		\begin{subfigure}{}
			\includegraphics[width=0.98\textwidth]{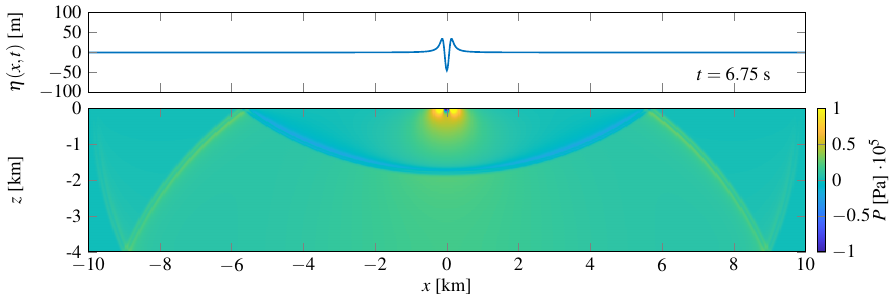}
		\end{subfigure}\\
		(c)\hfill
		\begin{subfigure}{}
			\includegraphics[width=0.98\textwidth]{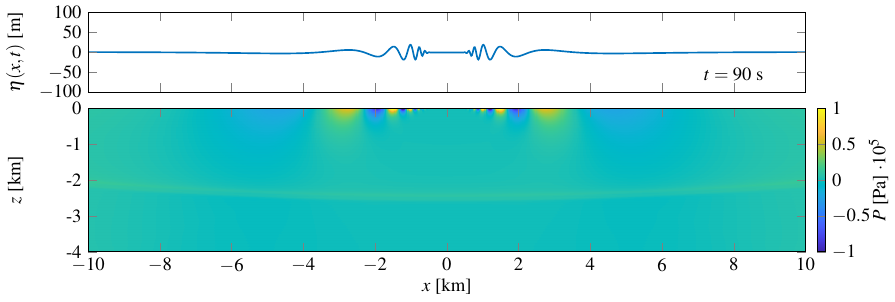}
		\end{subfigure}\\
		\caption{Compression waves caused by an initial Gaussian pressure distribution centred at $(0,0)$, including the effects of static compression.
		}
		\label{fig:surface_gaussian_static}
	\end{figure}

\begin{figure}[!h]
		\centering    
		(a)\hfill
		\begin{subfigure}{}
			\includegraphics[width=0.98\textwidth]{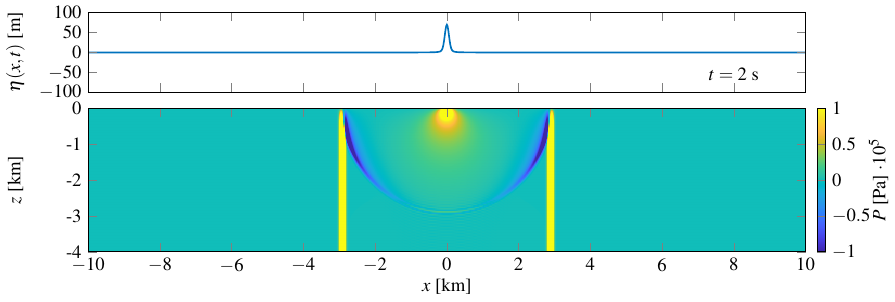}
		\end{subfigure}\\
		(b)\hfill
		\begin{subfigure}{}
			\includegraphics[width=0.98\textwidth]{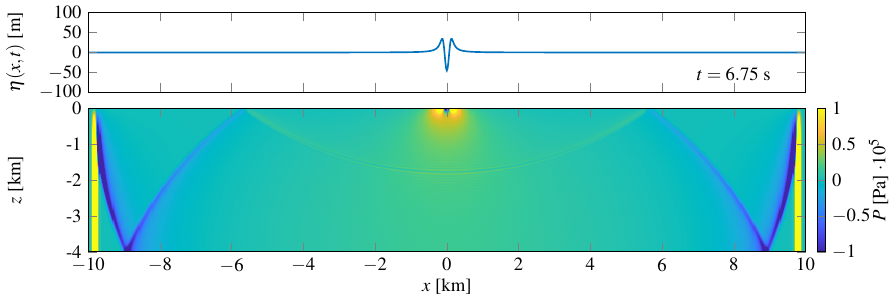}
		\end{subfigure}\\
		(c)\hfill
		\begin{subfigure}{}
			\includegraphics[width=0.98\textwidth]{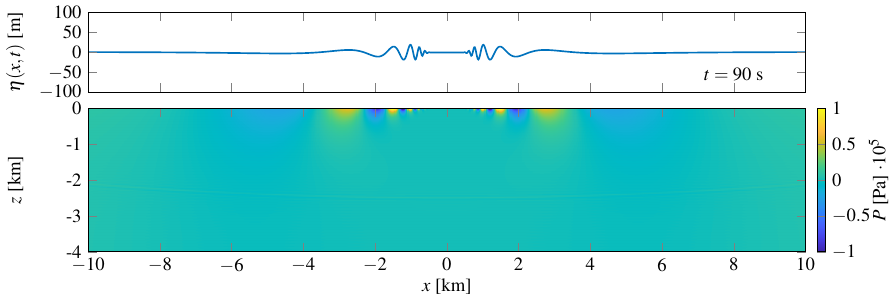}
		\end{subfigure}\\
		\caption{Compression waves caused by an initial one-dimensional Gaussian pressure distribution centred at $x_c=0$, including the effects of static compression.
		}
		\label{fig:surface_gaussian_static2}
	\end{figure}

	Comparing the results for the localised Gaussian (Figure \ref{fig:surface_gaussian_static}) with those for the one-dimensional Gaussian (Figure \ref{fig:surface_gaussian_static2}), we can see that the initial pressures excite propagating gravity waves on the surface of the same shape, which move much slower than the pressure waves. These gravity waves are orders of magnitude greater than the surface displacement due to compression waves. This means that in the case that gravity waves and pressure waves are created by the same source, it would be hard to detect the effects of compression and further determine the initial pressure profile from surface observations. The pressure due to the gravity waves decays away from the surface, and therefore, underwater pressure measurements are better suited to observe the internal compression waves.
	
	Focusing now on the internal pressure waves for our two initial pressures, as can be expected, the radically different initial pressure distributions lead to different internal pressure behaviour. The pressure wave generated by the localised Gaussian (Figure \ref{fig:surface_gaussian_static}) propagates radially, reflecting off the bottom and surface, creating a similar surface knocking phenomenon to the submerged Gaussian. The one-dimensional Gaussian (Figure \ref{fig:surface_gaussian_static2}), on the other hand, predominately propagates in the $x$-direction with a rapidly decaying trail that reflects off the surface, which does not produce an observable knocking behaviour. The respective animations are provided in Movie $4$ and Movie $5$.

	
	\section{Conclusion}\label{sec:conclusion}
	Under the assumptions of linearised water wave theory, this paper has considered the IBVP governing the evolution of a pressure disturbance in a two-dimensional ocean with a free surface, including the effect of compression. The problem was solved mathematically using self-adjoint operator theory by introducing an appropriate inner product. We presented two versions of the theory; in the first, we ignored the effect of static compression and included only the dynamic compression, while the second included both dynamic and static compression. Time-domain simulations showed the propagation of the pressure wave away from the initial source, reflecting off the free surface and the rigid ocean floor, and the onset of surface gravity waves propagating away from the source. In particular, a near reversal of phase is observed upon reflection from the surface, while the opposite pattern occurs in the reflection process from the rigid ocean floor. This indicates the fact that the free surface is approximately a homogeneous Dirichlet boundary. The corresponding time-domain simulations for the model incorporating static compression show that the effect of static compression is minor. While the present work deals with the canonical problem of uniform water depth, the theory considered here can be extended to tackle more realistic variable bathymetry using step-approximation, and it is equally applicable in modelling three-dimensional waves, albeit with increased computational requirements.

	\section*{Supplementary material}
	\noindent Animations for Figures \ref{fig:underwater_gaussian}--\ref{fig:surface_gaussian_static2}  are uploaded as supplementary material.
	%\section*{Acknowledgement}

	\subsection*{Funding}
	%\vspace{-0.3cm} 
	\noindent BW and MHM are supported by the Australian Research Council's Discovery Projects funding scheme (DP240102104).
	
	\subsection*{Declaration of interests}
	\noindent The authors report no conflict of interest.
	
	\subsection*{Data Availability}
	\noindent Data sharing does not apply to this article, as no new data was created or analysed in this study. The computer code used to generate the solutions is available from the corresponding author upon reasonable request.
	
	% \bibliographystyle{jfm}
	% \bibliography{References}  

\begin{thebibliography}{55}
		\expandafter\ifx\csname natexlab\endcsname\relax\def\natexlab#1{#1}\fi
		\def\au#1{#1} \def\ed#1{#1} \def\yr#1{#1}\def\at#1{#1}\def\jt#1{\textit{#1}} \def\bt#1{#1}\def\bvol#1{\textbf{#1}} \def\vol#1{#1} \def\pg#1{#1} \def\publ#1{#1}\def\arxiv#1{#1}\def\org#1{#1}\def\st#1{\textit{#1}}
		
		\bibitem[Abdolali {\em et~al.\/}(2019)Abdolali, Kadri \& Kirby]{abdolali2019effect}
		{\sc \au{Abdolali, A.}, \au{Kadri, U.} \& \au{Kirby, J.~T.}} \yr{2019}  \at{Effect of water compressibility, sea-floor elasticity, and field gravitational potential on tsunami phase speed}.  \jt{Sci. Rep.}  \bvol{9}~(1),  \pg{1--8}.
		
		\bibitem[Abdolali \& Kirby(2017)]{abdolali2017role}
		{\sc \au{Abdolali, A.} \& \au{Kirby, J.~T.}} \yr{2017}  \at{Role of compressibility on tsunami propagation}.  \jt{J. Geophys. Res. Oceans}  \bvol{122}~(12),  \pg{9780--9794}.
		
		\bibitem[Abdolali {\em et~al.\/}(2015)Abdolali, Kirby \& Bellotti]{abdolali2015depth}
		{\sc \au{Abdolali, A.}, \au{Kirby, J.~T.} \& \au{Bellotti, G.}} \yr{2015}  \at{Depth-integrated equation for hydro-acoustic waves with bottom damping}.  \jt{J. Fluid Mech.}  \bvol{766},  \pg{R1}.
		
		\bibitem[Andrade {\em et~al.\/}(2025)Andrade, Bustamante, Kadri \& Stuhlmeier]{andrade2025acoustic}
		{\sc \au{Andrade, D.}, \au{Bustamante, M.~D.}, \au{Kadri, U.} \& \au{Stuhlmeier, R.}} \yr{2025}  \at{Acoustic-gravity wave triad resonance in compressible flow: a dynamical systems approach}.  \jt{J. Fluid Mech.}  \bvol{1013},  \pg{A23}.
		
		\bibitem[Bondi(1947)]{bondi1947waves}
		{\sc \au{Bondi, H}} \yr{1947} Waves on the surface of a compressible liquid.  \bt{In {\em Math. Proc. Camb. Philos. Soc.\/}}, ,  \vol{vol.~43},  \pg{pp. 75--95}. Cambridge University Press.
		
		\bibitem[Chen {\em et~al.\/}(2019)Chen, Gilbert \& Guyenne]{chen2019dispersion}
		{\sc \au{Chen, H.}, \au{Gilbert, R.~P.} \& \au{Guyenne, P.}} \yr{2019}  \at{Dispersion and attenuation in a porous viscoelastic model for gravity waves on an ice-covered ocean}.  \jt{Eur. J. Mech. B-Fluid}  \bvol{78},  \pg{88--105}.
		
		\bibitem[Chierici {\em et~al.\/}(2010)Chierici, Pignagnoli \& Embriaco]{chierici2010modeling}
		{\sc \au{Chierici, F.}, \au{Pignagnoli, L.} \& \au{Embriaco, D.}} \yr{2010}  \at{Modeling of the hydroacoustic signal and tsunami wave generated by seafloor motion including a porous seabed}.  \jt{J. Geophys. Res-Oceans}  \bvol{115}~(C3),  \pg{C03015}.
		
		\bibitem[Dahlquist \& Björck(2008)]{Dahlquist_Björck_2008}
		{\sc \au{Dahlquist, G.} \& \au{Björck, Å.}} \yr{2008} {\em Numerical methods in scientific computing, volume I\/}.  \publ{SIAM}.
		
		\bibitem[Das \& Meylan(2023{\natexlab{{\em a\/}}})]{Das2023PoF}
		{\sc \au{Das, S.} \& \au{Meylan, M.~H.}} \yr{2023{\natexlab{{\em a\/}}}}  \at{Effect of static compression on tsunami waves: Two-dimensional solution}.  \jt{Phys. Fluids}  \bvol{36},  \pg{066603}.
		
		\bibitem[Das \& Meylan(2023{\natexlab{{\em b\/}}})]{Das2023time-domain}
		{\sc \au{Das, S.} \& \au{Meylan, M.~H.}} \yr{2023{\natexlab{{\em b\/}}}}  \at{Time-domain wave response of a compressible ocean due to an arbitrary ocean bottom motion}.  \jt{Appl. Math. Model.}  \bvol{118},  \pg{832--852}.
		
		\bibitem[Das \& Meylan(2024)]{das4636389effect}
		{\sc \au{Das, S.} \& \au{Meylan, M.~H.}} \yr{2024}  \at{Effect of static compression on near-field tsunami waves: Three-dimensional solution}.  \jt{Eur. j. Mech. B Fluids}  \bvol{106},  \pg{197--213}.
		
		\bibitem[Das {\em et~al.\/}(2024)Das, Pethiyagoda \& Meylan]{Das2024}
		{\sc \au{Das, S.}, \au{Pethiyagoda, R.} \& \au{Meylan, M.~H.}} \yr{2024}  \at{Compressible ocean waves generated by sudden seabed rise near a step-type topography}.  \jt{Phys. Fluids}  \bvol{36},  \pg{106619}.
		
		\bibitem[Dubois {\em et~al.\/}(2023)Dubois, Imperiale, Mangeney, Bouchut \& Sainte-Marie]{dubois2023acoustic}
		{\sc \au{Dubois, J.}, \au{Imperiale, S.}, \au{Mangeney, A.}, \au{Bouchut, F.} \& \au{Sainte-Marie, J.}} \yr{2023}  \at{Acoustic and gravity waves in the ocean: a new derivation of a linear model from the compressible euler equation}.  \jt{J. Fluid Mech.}  \bvol{970},  \pg{A28}.
		
		\bibitem[Eyov {\em et~al.\/}(2013)Eyov, Klar, Kadri \& Stiassnie]{eyov2013progressive}
		{\sc \au{Eyov, E.}, \au{Klar, A.}, \au{Kadri, U.} \& \au{Stiassnie, M.}} \yr{2013}  \at{Progressive waves in a compressible-ocean with an elastic bottom}.  \jt{Wave Motion}  \bvol{50}~(5),  \pg{929--939}.
		
		\bibitem[Filon(1930)]{filon_quadrature_1930}
		{\sc \au{Filon, L. N.~G.}} \yr{1930}  \at{{III}.—{On} a {Quadrature} {Formula} for {Trigonometric} {Integrals}}.  \jt{Proc. Roy. Soc. Edinburgh}  \bvol{49},  \pg{38--47}.
		
		\bibitem[Hazard \& Loret(2007)]{hazard2007generalized}
		{\sc \au{Hazard, C.} \& \au{Loret, F.}} \yr{2007}  \at{Generalized eigenfunction expansions for conservative scattering problems with an application to water waves}.  \jt{Proc. Roy. Soc. Edinburgh A}  \bvol{137}~(5),  \pg{995--1035}.
		
		\bibitem[Hazard \& Meylan(2008)]{hazard2008spectral}
		{\sc \au{Hazard, C.} \& \au{Meylan, M.~H.}} \yr{2008}  \at{Spectral theory for an elastic thin plate floating on water of finite depth}.  \jt{SIAM J. Appl. Math.}  \bvol{68}~(3),  \pg{629--647}.
		
		\bibitem[Hendin \& Stiassnie(2013)]{hendin2013tsunami}
		{\sc \au{Hendin, G.} \& \au{Stiassnie, M.}} \yr{2013}  \at{Tsunami and acoustic-gravity waves in water of constant depth}.  \jt{Phys. Fluids}  \bvol{25}~(8),  \pg{086103}.
		
		\bibitem[Ikebe(1960)]{ikebe1960eigenfunction}
		{\sc \au{Ikebe, T.}} \yr{1960}  \at{Eigenfunction expansions associated with the {S}chroedinger operators and their applications to scattering theory}.  \jt{Arch. Ration. Mech. Anal.}  \bvol{5},  \pg{1--34}.
		
		\bibitem[Kadri(2014)]{kadri2014deep}
		{\sc \au{Kadri, U.}} \yr{2014}  \at{Deep ocean water transport by acoustic-gravity waves}.  \jt{J. Geophys. Res- Ocean}  \bvol{119}~(11),  \pg{7925--7930}.
		
		\bibitem[Kadri(2015)]{kadri2015wave}
		{\sc \au{Kadri, U.}} \yr{2015}  \at{Wave motion in a heavy compressible fluid: Revisited}.  \jt{Eur. J. Mech. B-Fluid.}  \bvol{49},  \pg{50--57}.
		
		\bibitem[Kadri(2016)]{kadri2016triad}
		{\sc \au{Kadri, U.}} \yr{2016}  \at{Triad resonance between a surface-gravity wave and two high frequency hydro-acoustic waves}.  \jt{Eur. J. Mech. B-Fluid.}  \bvol{55},  \pg{157--161}.
		
		\bibitem[Kadri(2019)]{kadri2019effect}
		{\sc \au{Kadri, U.}} \yr{2019}  \at{Effect of sea-bottom elasticity on the propagation of acoustic--gravity waves from impacting objects}.  \jt{Sci. Rep.}  \bvol{9}~(1),  \pg{912}.
		
		\bibitem[Kadri(2024)]{kadri2024underwater}
		{\sc \au{Kadri, U.}} \yr{2024}  \at{Underwater acoustic analysis reveals unique pressure signals associated with aircraft crashes in the sea: revisiting {MH}370}.  \jt{Sci. Rep.}  \bvol{14}~(1),  \pg{10102}.
		
		\bibitem[Kadri \& Akylas(2016)]{kadri2016resonant}
		{\sc \au{Kadri, U.} \& \au{Akylas, T.~R.}} \yr{2016}  \at{On resonant triad interactions of acoustic--gravity waves}.  \jt{J. Fluid Mech.}  \bvol{788},  \pg{R1}.
		
		\bibitem[Kadri \& Stiassnie(2012)]{kadri2012acoustic}
		{\sc \au{Kadri, U.} \& \au{Stiassnie, M.}} \yr{2012}  \at{Acoustic-gravity waves interacting with the shelf break}.  \jt{J. Geophys. Res- Ocean}  \bvol{117}~(C3),  \pg{C03035}.
		
		\bibitem[Kadri \& Stiassnie(2013)]{kadri2013generation}
		{\sc \au{Kadri, U.} \& \au{Stiassnie, M.}} \yr{2013}  \at{Generation of an acoustic-gravity wave by two gravity waves, and their subsequent mutual interaction}.  \jt{J. Fluid Mech.}  \bvol{735},  \pg{R6}.
		
		\bibitem[Lamb(1957)]{lamb1924hydrodynamics}
		{\sc \au{Lamb, Horace}} \yr{1957} {\em Hydrodynamics\/}, 6th edn.  \publ{University Press}.
		
		\bibitem[Liu \& Higuera(2022)]{liu2022water}
		{\sc \au{Liu, P. L-F.} \& \au{Higuera, P.}} \yr{2022}  \at{Water waves generated by moving atmospheric pressure: theoretical analyses with applications to the 2022 {Tonga} event}.  \jt{J. Fluid Mech.}  \bvol{951},  \pg{A34}.
		
		\bibitem[Longuet-Higgins(1950)]{longuet1950theory}
		{\sc \au{Longuet-Higgins, M.~S.}} \yr{1950}  \at{A theory of the origin of microseisms}.  \jt{Philos. Tr. R. Soc. S-A}  \bvol{243}~(857),  \pg{1--35}.
		
		\bibitem[Meylan(2002)]{meylan_2002}
		{\sc \au{Meylan, M.~H.}} \yr{2002}  \at{Spectral solution of time-dependent shallow water hydroelasticity}.  \jt{J. Fluid Mech.}  \bvol{454},  \pg{387–402}.
		
		\bibitem[Meylan(2009)]{meylan_2009}
		{\sc \au{Meylan, M.~H.}} \yr{2009}  \at{Time-dependent linear water-wave scattering in two dimensions by a generalized eigenfunction expansion}.  \jt{J. Fluid Mech.}  \bvol{632},  \pg{447–455}.
		
		\bibitem[Meylan \& Eatock~Taylor(2009)]{meylan2009time2}
		{\sc \au{Meylan, M.~H.} \& \au{Eatock~Taylor, R.}} \yr{2009}  \at{Time-dependent water-wave scattering by arrays of cylinders and the approximation of near trapping}.  \jt{J. Fluid Mech.}  \bvol{631},  \pg{103--125}.
		
		\bibitem[Michele \& Renzi(2020)]{michele2020effects}
		{\sc \au{Michele, S.} \& \au{Renzi, E.}} \yr{2020}  \at{Effects of the sound speed vertical profile on the evolution of hydroacoustic waves}.  \jt{J. Fluid Mech.}  \bvol{883},  \pg{A28}.
		
		\bibitem[Miyoshi(1954)]{miyoshi1954generation}
		{\sc \au{Miyoshi, H.}} \yr{1954}  \at{Generation of the tsunami in compressible water ({P}art {I})}.  \jt{J. Oceanogr. Soc. Japan}  \bvol{10}~(1),  \pg{1--9}.
		
		\bibitem[Nosov {\em et~al.\/}(2018)Nosov, Karpov, Kolesov, Sementsov, Matsumoto \& Kaneda]{nosov2018relationship}
		{\sc \au{Nosov, M.}, \au{Karpov, V.}, \au{Kolesov, S.}, \au{Sementsov, K.}, \au{Matsumoto, H.} \& \au{Kaneda, Y.}} \yr{2018}  \at{Relationship between pressure variations at the ocean bottom and the acceleration of its motion during a submarine earthquake}.  \jt{Earth Planet. Space}  \bvol{70}~(1),  \pg{100}.
		
		\bibitem[Nosov(1999)]{nosov1999tsunami}
		{\sc \au{Nosov, M.~A.}} \yr{1999}  \at{Tsunami generation in compressible ocean}.  \jt{Phys. Chem. Earth Pt. B}  \bvol{24}~(5),  \pg{437--441}.
		
		\bibitem[Nosov \& Kolesov(2007)]{nosov2007elastic}
		{\sc \au{Nosov, M.~A.} \& \au{Kolesov, S.~V.}} \yr{2007}  \at{Elastic oscillations of water column in the 2003 {T}okachi-oki tsunami source: in-situ measurements and 3-{D} numerical modelling}.  \jt{Nat. Hazard. Earth Sys.}  \bvol{7}~(3),  \pg{243--249}.
		
		\bibitem[Nosov \& Skachko(2001)]{nosov2001nonlinear}
		{\sc \au{Nosov, M.~A.} \& \au{Skachko, S.~N.}} \yr{2001}  \at{Nonlinear tsunami generation mechanism}.  \jt{Nat. Hazard. Earth Sys.}  \bvol{1},  \pg{251--253}.
		
		\bibitem[Omira {\em et~al.\/}(2022)Omira, Ramalho, Kim, Gonz{\'a}lez, Kadri, Miranda, Carrilho \& Baptista]{omira2022global}
		{\sc \au{Omira, R.}, \au{Ramalho, R.~S.}, \au{Kim, J.}, \au{Gonz{\'a}lez, P.~J.}, \au{Kadri, U.}, \au{Miranda, J.~M.}, \au{Carrilho, F.} \& \au{Baptista, M.~A.}} \yr{2022}  \at{Global {T}onga tsunami explained by a fast-moving atmospheric source}.  \jt{Nature}  \bvol{609}~(7928),  \pg{734--740}.
		
		\bibitem[Peter \& Meylan(2010)]{peter2010general}
		{\sc \au{Peter, M.~A.} \& \au{Meylan, M.~H.}} \yr{2010}  \at{A general spectral approach to the time-domain evolution of linear water waves impacting on a vertical elastic plate}.  \jt{SIAM J. Appl. Math.}  \bvol{70}~(7),  \pg{2308--2328}.
		
		\bibitem[Pethiyagoda {\em et~al.\/}(2024{\natexlab{{\em a\/}}})Pethiyagoda, Das, Bonham \& Meylan]{pethiyagoda_3d_atmos}
		{\sc \au{Pethiyagoda, R.}, \au{Das, S.}, \au{Bonham, M.} \& \au{Meylan, M.~H.}} \yr{2024{\natexlab{{\em a\/}}}}  \at{Atmospheric pressure-induced three-dimensional surface wave propagation in the compressible ocean: Effect of static compression}.  \jt{Phys. Fluids}  \bvol{36},  \pg{096618}.
		
		\bibitem[Pethiyagoda {\em et~al.\/}(2024{\natexlab{{\em b\/}}})Pethiyagoda, Das \& Meylan]{Pethiyagoda_Das_Meylan_2024}
		{\sc \au{Pethiyagoda, R.}, \au{Das, S.} \& \au{Meylan, M.~H.}} \yr{2024{\natexlab{{\em b\/}}}}  \at{Evolution of arbitrary temporal ocean floor motion-induced surface waves over a trench}.  \jt{Phys. Fluids}  \bvol{36},  \pg{126606}.
		
		\bibitem[Pethiyagoda {\em et~al.\/}(2025{\natexlab{{\em a\/}}})Pethiyagoda, Das \& Meylan]{pethiyagoda4677642atmospheric}
		{\sc \au{Pethiyagoda, R.}, \au{Das, S.} \& \au{Meylan, M.~H.}} \yr{2025{\natexlab{{\em a\/}}}}  \at{Atmospheric pressure-driven surface wave propagation in a compressible ocean including static compression}.  \jt{Wave Motion}  \bvol{134},  \pg{103468}.
		
		\bibitem[Pethiyagoda {\em et~al.\/}(2025{\natexlab{{\em b\/}}})Pethiyagoda, Das \& Meylan]{pethiyagoda_arbitrary25}
		{\sc \au{Pethiyagoda, R.}, \au{Das, S.} \& \au{Meylan, M.~H}} \yr{2025{\natexlab{{\em b\/}}}}  \at{Time-domain pressure and surface wave propagation over generic topography due to sea floor motion}.  \jt{arXiv preprint arXiv:2509.23377} .
		
		\bibitem[Povzner(1953)]{povzner1953expansion}
		{\sc \au{Povzner, A.~Y.}} \yr{1953}  \at{On the expansion of arbitrary functions in characteristic functions of the operator -{$\Delta u+cu$} (in russian)}.  \jt{Matematicheskii Sbornik}  \bvol{74}~(1),  \pg{109--156}.
		
		\bibitem[Renzi \& Dias(2014)]{renzi2014hydro}
		{\sc \au{Renzi, E.} \& \au{Dias, F.}} \yr{2014}  \at{Hydro-acoustic precursors of gravity waves generated by surface pressure disturbances localised in space and time}.  \jt{J. Fluid Mech.}  \bvol{754},  \pg{250--262}.
		
		\bibitem[Sammarco {\em et~al.\/}(2013)Sammarco, Cecioni, Bellotti \& Abdolali]{sammarco2013depth}
		{\sc \au{Sammarco, P.}, \au{Cecioni, C.}, \au{Bellotti, G.} \& \au{Abdolali, A.}} \yr{2013}  \at{Depth-integrated equation for large-scale modelling of low-frequency hydroacoustic waves}.  \jt{J. Fluid Mech.}  \bvol{722},  \pg{R6}.
		
		\bibitem[Sells(1965)]{sells1965effect}
		{\sc \au{Sells, C. C.~L.}} \yr{1965}  \at{The effect of a sudden change of shape of the bottom of a slightly compressible ocean}.  \jt{Philos. Tr. R. Soc. Ser. A}  \bvol{258}~(1092),  \pg{495--528}.
		
		\bibitem[Stiassnie(2010)]{stiassnie2010tsunamis}
		{\sc \au{Stiassnie, M.}} \yr{2010}  \at{Tsunamis and acoustic-gravity waves from underwater earthquakes}.  \jt{J. Eng. Math.}  \bvol{67}~(1-2),  \pg{23--32}.
		
		\bibitem[Wilcox(1975)]{wilcox1975scattering}
		{\sc \au{Wilcox, C.~H.}} \yr{1975} {\em Scattering theory for the d'Alembert equation in exterior domains\/}, ,  \vol{vol. 442}.  \publ{Springer-Verlag}.
		
		\bibitem[Wilks {\em et~al.\/}(2024)Wilks, Meylan, Montiel \& Wakes]{wilks2024}
		{\sc \au{Wilks, B.}, \au{Meylan, M.~H.}, \au{Montiel, F.} \& \au{Wakes, S.}} \yr{2024}  \at{Generalized eigenfunction expansion and singularity expansion methods for two-dimensional acoustic time-domain wave scattering problems}.  \jt{Proc. R. Soc, A: Math. Phys. Eng. Sci.}  \bvol{480}~(2297),  \pg{20230845}.
		
		\bibitem[Wilks {\em et~al.\/}(2025)Wilks, Meylan, Montiel \& Wakes]{WILKS2025103421}
		{\sc \au{Wilks, B.}, \au{Meylan, M.~H.}, \au{Montiel, F.} \& \au{Wakes, S.}} \yr{2025}  \at{Generalised eigenfunction expansion and singularity expansion methods for canonical time-domain wave scattering problems}.  \jt{Wave Motion}  \bvol{132},  \pg{103421}.
		
		\bibitem[Williams \& Kadri(2023)]{williams2023acoustic}
		{\sc \au{Williams, B.} \& \au{Kadri, U.}} \yr{2023}  \at{On the propagation of acoustic–gravity waves due to a slender rupture in an elastic seabed}.  \jt{J. Fluid Mech.}  \bvol{956}.
		
		\bibitem[Yamamoto(1982)]{yamamoto1982gravity}
		{\sc \au{Yamamoto, T.}} \yr{1982}  \at{Gravity waves and acoustic waves generated by submarine earthquakes}.  \jt{Int. J. Soil Dyn. Earthq. Eng.}  \bvol{1}~(2),  \pg{75--82}.
		
	\end{thebibliography}
	
	\providecommand{\noopsort}[1]{}\providecommand{\singleletter}[1]{#1}%

	\appendix
	\section{Orthogonality of the eigenfunctions}\label{sec:orthogonality}
	Here, we determine the normalisation coefficients $D_p(k)$ while simultaneously confirming the orthogonality of the eigenfunctions. Firstly, we calculate the $L^2(-h,0)$ inner product of the depth-dependent functions through two applications of integration by parts, yielding
	\begin{subequations}
		\allowdisplaybreaks
		\begin{align}
			&\int_{-h}^0f_n(z;k)f_p(z;\tilde{k})\,{\rm d}z=\int_{-h}^0 \frac{\displaystyle \cosh{\mu_n(z+h)}}{\cosh{\mu_n h}}\frac{\displaystyle \cosh{\tilde{\mu}_p(z+h)}}{\cosh{\tilde{\mu}_p h}}\,{\rm d}z\nonumber\\
			&\hspace{1.4in} =\frac{1}{\mu_n}\tanh{\mu_n h}-\frac{\tilde{\mu}_p}{\mu_n^2}\tanh{\tilde{\mu}_p h}+\frac{\tilde{\mu}_p^2}{\mu_n^2}\int_{-h}^0f_n(z;k)f_p(z;\tilde{k})\,{\rm d}z,\nonumber\\
		\end{align}
		where for shorthand we write $\mu_n=\mu_n(k)$ and $\tilde{\mu_p}=\mu_p(\tilde{k})$. Rearranging using (\ref{eq:dispersion_relation}) and (\ref{eq:dispersion_relation2}) to remove instances of $\mathrm{tanh}(\cdot)$ and $k$ gives
		\begin{align}
			\int_{-h}^0f_n(z;k)f_p(z;\tilde{k})\,{\rm d}z=\frac{1}{g}\frac{\omega_n^2-\tilde{\omega}_p^2}{\mu_n^2-\tilde{\mu}_p^2},
		\end{align}
	\end{subequations}
	with $\omega_n=\omega_n(k)$ and $\tilde{\omega_p}=\omega_p(\tilde{k})$.
	Now we compute the inner product $\langle\phi_n^\pm,\phi_p^\pm\rangle_{\mathcal{E}}$ as
	\begin{subequations}\label{eq:inner_product}
		\begin{align}
			\langle\phi_n^\pm(\cdot,\cdot,k),\phi_p^\pm(\cdot,\cdot,\tilde{k}) \rangle_{\mathcal{E}} &=\int_{-\infty}^\infty\int_{-h}^0 \varphi_n^\pm(x,z;k)\overline{\varphi_p^\pm(x,z;\tilde{k})}\,{\rm d}z\,{\rm d}x\nonumber\\
			&\quad\quad+\frac{c^2}{g}\int_{-\infty}^\infty\varphi_n^\pm(x,0;k) \overline{\varphi_p^\pm(x,0;\tilde{k})}\,{\rm d}x,\nonumber\\
			&=\int_{-\infty}^\infty\left(\frac{\omega_n^2-\tilde{\omega}_p^2}{\mu_n^2-\tilde{\mu}_p^2}+\frac{c^2}{g}\right)e^{\mp{\rm i}(k-\tilde{k}) x}\,{\rm d}x,\nonumber\\
			&=\frac{c^2}{g}\int_{-\infty}^\infty\left(\frac{k^2-\tilde{k}^2}{\mu_n^2-\tilde{\mu}_p^2}\right)e^{\mp{\rm i}(k-\tilde{k}) x}\,{\rm d}x,\nonumber\\
			&=2\pi\frac{c^2}{g}\left(\frac{k^2-\tilde{k}^2}{\mu_n^2-\tilde{\mu}_p^2}\right)\delta(k-\tilde{k}),\label{eq:inner_product_same}
		\end{align}
		where $\delta(\kappa)$ is the Dirac delta. Finally, in the case of $n=p$, we can recover the normalisation constant by taking limits,
		\[\lim_{k\rightarrow\tilde{k}}\langle \phi_n^\pm(\cdot,\cdot,k),\phi_p^\pm(\cdot,\cdot,\tilde{k})\rangle_{\mathcal{E}}= 2\pi\frac{c^2}{g}\left(\frac{\tilde{k}}{\mu_n\partial_k\tilde{\mu}_n}\right)\delta(k-\tilde{k})\delta_{np},\]
		where
		\[
		\partial_k \mu_n=\frac{2\dfrac{c^2}{g}k}{\mu_n h{\rm sech}^2{\mu_n h}+\tanh{\mu_n h}+2\dfrac{c^2}{g}\mu_n}.
		\]
		For simplicity, we write
		\[\lim_{k\rightarrow\tilde{k}}\langle \phi_n^\pm(\cdot,\cdot,k),\phi_p^\pm(\cdot,\cdot,\tilde{k})\rangle_{\mathcal{E}}= D_p(k)\delta(k-\tilde{k})\delta_{np},\]
		where
		\[
		D_p(k)=2\pi\left(\frac{h}{2}{\rm sech}^2{\mu_p h}+\frac{\tanh{\mu_p h}}{2\mu_p}+\frac{c^2}{g}\right).
		\]
		Similarly,
		\begin{align}
			\langle\phi_n^\pm,\phi_p^\mp \rangle_{\mathcal{E}} =2\pi\frac{c^2}{g}\left(\frac{k^2-\tilde{k}^2}{\mu_n^2-\tilde{\mu}_p^2}\right)\delta(k+\tilde{k})=0,\label{eq:inner_product_opposite}
		\end{align}
	\end{subequations}
	since $k$ and $\tilde{k}$ are both positive real numbers.
	
	\section{Filon-type quadrature}\label{sec:filon-type}
	Following the method of \citet{filon_quadrature_1930}, we compute an integral of the form
	\[
	I=\int_{a}^{b}f(x)e^{{\rm i}\psi(x)}\,dx,
	\]
	where $f(x)$ and $\psi'(x)$ are slowly varying functions, by splitting the domain of integration at the points $x_0=a,x_1,\ldots,x_N=b$ and approximating the functions $f(x)$ and $\psi(x)$ by piecewise linear functions
	\allowdisplaybreaks
	\begin{align*}
		I&=\sum_{i=0}^{N-1}\int_{x_i}^{x_{i+1}}\left(f_i+\frac{f_{i+1}-f_i}{\Delta x_i}(x-x_i)\right)e^{{\rm i}\left(\psi_i+\frac{\psi_{i+1}-\psi_i}{\Delta x_i}(x-x_i)\right)}dx,\\
		&=\sum_{i=0}^{N-1}\Delta x_i\int_{0}^{1}\left(f_i+(f_{i+1}-f_i)s\right)e^{{\rm i}\left(\psi_i+(\psi_{i+1}-\psi_i)s\right)}ds,\quad\mbox{where }s=\frac{x-x_i}{\Delta x_i},\\
		&=\sum_{i=0}^{N-1}\Delta x_i\left[-\frac{{\rm i}}{\psi_{i+1}-\psi_i}\left(f_{i+1}e^{{\rm i}\psi_{i+1}}
		-f_ie^{{\rm i}\psi_i}\right)
		+\frac{f_{i+1}-f_i}{(\psi_{i+1}-\psi_i)^2}\left(e^{{\rm i}\psi_{i+1}}-e^{{\rm i}\psi_i}\right)\right],\\
		&=\sum_{i=0}^{N-1}\Delta x_i\left[\left(\frac{{\rm i}e^{{\rm i}\psi_i}}{\psi_{i+1}-\psi_i}-\frac{e^{{\rm i}\psi_{i+1}}-e^{{\rm i}\psi_i}}{(\psi_{i+1}-\psi_i)^2}\right)f_i
		+\left(-\frac{{\rm i}e^{{\rm i}\psi_{i+1}}}{\psi_{i+1}-\psi_i}+\frac{e^{{\rm i}\psi_{i+1}}-e^{{\rm i}\psi_i}}{(\psi_{i+1}-\psi_i)^2}\right)f_{i+1}\right],\\
		&=\Delta x_0\left(\frac{{\rm i}e^{{\rm i}\psi_0}}{\psi_{1}-\psi_0}-\frac{e^{{\rm i}\psi_{1}}-e^{{\rm i}\psi_0}}{(\psi_{1}-\psi_0)^2}\right)f_0+\sum_{i=1}^{N-1}\left[\Delta x_{i-1}\left(-\frac{{\rm i}e^{{\rm i}\psi_{i}}}{\psi_{i}-\psi_{i-1}}+\frac{e^{{\rm i}\psi_{i}}-e^{{\rm i}\psi_{i-1}}}{(\psi_{i}-\psi_{i-1})^2}\right)\right.\\
		&\quad\left.+\Delta x_i\left(\frac{{\rm i}e^{{\rm i}\psi_i}}{\psi_{i+1}-\psi_i}-\frac{e^{{\rm i}\psi_{i+1}}-e^{{\rm i}\psi_i}}{(\psi_{i+1}-\psi_i)^2}\right)\right]f_i+\Delta x_{N-1}\left(-\frac{{\rm i}e^{{\rm i}\psi_{N}}}{\psi_{N}-\psi_{N-1}}+\frac{e^{{\rm i}\psi_{N}}-e^{{\rm i}\psi_{N-1}}}{(\psi_{N}-\psi_{N-1})^2}\right)f_{N}.
	\end{align*}
	The result is a weighting scheme for quadrature integration that can account for a rapidly changing phase function.
	
	\section{Errors for numerical parameter choice}\label{app:error}
	This section compares the effect that the numerical scheme parameters $N$, $k_\mathrm{max}$, and $\Delta k$ have on the error in reconstructing the initial condition for the underwater Gaussian pressure disturbance (\ref{eq:inital_pressure}). The physical parameters used are $\hat{P}=10^6$ Pa, $(x_c,z_c)=(0,-2000)$ m, $g=9.81$m/s$^2$, $c=1450$m/s, $h=4000$m and $\sigma=200$m. The error measure is given by
	\[
	\max_{x,z}\frac{\rho\Phi_t(x,z,0)+P_0(x,y)}{\hat{P}},
	\]
	as the strength of the pressure $\hat{P}$ acts as a linear scaling of the solution.
	
	Tables \ref{tab:num_error_N}-\ref{tab:num_error_Nk} present the errors when varying $N$, $k_\mathrm{max}$, and $\Delta k$, respectively. When not varied, the default parameter values $N=100$, $k_\mathrm{max}=0.2$, and $\Delta k=0.0002$ are used. We note that the error given by the default parameters is sufficient for visual convergence. Of the three parameters, the number of included modes $N$ (Table \ref{tab:num_error_N}) has the most significant effect on the accuracy of the solution because it determines how well the initial condition can be resolved in the vertical direction. The choice of $k_\mathrm{max}$ (Table \ref{tab:num_error_kmax}) is affected by the choice of pressure distribution, in our case, leading to exponential decay in the integrand. Finally, all tested discretisation spacings $\Delta k$ (Table \ref{tab:num_error_Nk}) resulted in similar error, indicating that we have sufficient resolution of our integrand to numerically perform the integration.
	\begin{table}
		\centering
		\begin{tabular}{cc}
			$N$ & Error \\
			25 & 3.77$\times 10^{-1}$\\
			50 & 7.71$\times 10^{-2}$\\
			100 & 4.01$\times 10^{-4}$\\
			200 & 7.66$\times 10^{-6}$\\
		\end{tabular}
		\caption{Initial condition reconstruction error for different $N$, and $k_\mathrm{max}=0.2$ and $\Delta k=0.0002$.}
		\label{tab:num_error_N}
	\end{table}
	
	\begin{table}
		\centering
		\begin{tabular}{cc}
			$k_\mathrm{max}$ & Error \\
			0.05 & 2.48$\times 10^{-2}$\\
			0.1 & 4.09$\times 10^{-4}$\\
			0.2 & 4.01$\times 10^{-4}$\\
			0.4 & 4.02$\times 10^{-4}$\\
		\end{tabular}
		\caption{Initial condition reconstruction error for different $k_\mathrm{max}$, and $N=100$ and $\Delta k=0.0002$.}
		
		\label{tab:num_error_kmax}
	\end{table}
	\begin{table}
		\centering
		\begin{tabular}{cc}
			$\Delta k$ & Error \\
			0.0008 & 4.02$\times 10^{-4}$\\
			0.0004 & 4.01$\times 10^{-4}$\\
			0.0002 & 4.01$\times 10^{-4}$\\
			0.0001 & 4.01$\times 10^{-4}$\\
		\end{tabular}
		\caption{Initial condition reconstruction error for different $\Delta k$, and $k_\mathrm{max}=0.2$ and $N=100$.}
		\label{tab:num_error_Nk}
	\end{table}
	
	\section{Normalisation coefficients in the case of static compression}\label{normalisation_appendix}
	In this section, we compute the normalisation coefficients $D_p(k)$ needed in the solution in the case of static compression, as discussed in \textsection\ref{static_compression_sec}. We begin by introducing a new inner product over the vertical ordinate
	\begin{equation}\label{IP_stat}
		\langle u_1,u_2\rangle_h=\int_{-h}^0 e^{-\gamma z}u_1(z)\overline{u_2(z)}\,{\rm d}z+\frac{c^2}{g}u_1(0)\overline{u_2(0)},
	\end{equation}
	and seek to compute $\langle f_n(\cdot;k),f_m(\cdot;\tilde{k})\rangle_h$. Following some tedious calculations, we find
	\begin{subequations}
		\begin{align}
			&\left(\mu_n^2-\frac{\gamma^2}{4}\right)\int_{-h}^0 e^{-\gamma z}f_n(z;k)f_m(z;\tilde{k})\,{\rm d}z=\left[e^{-\gamma z}f_n'(z;k)f_m(z;\tilde{k})\right]_{-h}^0-\left[e^{-\gamma z}f_n(z;k)f_m'(z;\tilde{k})\right]_{-h}^0\nonumber\\
			&\hspace{2in}+\int_{-h}^0 e^{-\gamma z}f_n(z;k)\left(\tilde{\mu}_m^2-\frac{\gamma^2}{4}\right)f_m(z;\tilde{k})\,{\rm d}z,   
		\end{align}
		which yields
		\begin{align}
			\int_{-h}^0 e^{-\gamma z}f_n(z;k)f_m(z;\tilde{k})\,{\rm d}z=\frac{\omega_n^2(k)-\omega_m^2(\tilde{k})}{g \left(\mu_n^2-\tilde{\mu}_m^2\right)}.
		\end{align}
	\end{subequations}
	Now, using \eqref{IP_stat},
	\begin{subequations}
		\allowdisplaybreaks
		\begin{align}
			\langle f_n(z;k),f_m(z;\tilde{k})\rangle_h&=\frac{\omega_n^2(k)-\omega_m^2(\tilde{k})}{g \left(\mu_n^2-\tilde{\mu}_m^2\right)}f_n(0;k)f_m(0;\tilde{k})+\frac{c^2}{g},\nonumber\\
			&=\frac{c^2}{g}\frac{k^2-\tilde{k}^2}{\mu_n^2-\tilde{\mu}_m^2}.
		\end{align}
		When $n=m$, the above inner product reduces to
		\begin{align}
			\lim_{k\rightarrow\tilde{k}} \langle f_m(z;k),f_m(z;\tilde{k})\rangle_h = \frac{c^2}{g}\frac{\tilde{k}}{\tilde{\mu}_m\partial_k\tilde{\mu}_m},
		\end{align}
		where
		\begin{align}
			&\partial_k\mu_n=\frac{\displaystyle 2\frac{c^2}{g}\left(1-\frac{\gamma}{2}\frac{\tanh{\mu_n h}}{\mu_n}\right)k}{\mathcal{A}(\mu_n,k)\tanh{\mu_n h} + \mathcal{B}(\mu_n,k)\mathrm{sech}^2{\mu_n h}+2\dfrac{c^2}{g}\mu_n},\\
			\mbox{with }& \mathcal{A}(\mu_n,k)= \left(1-\frac{\gamma c^2}{2g}\right) +\left( \frac{\gamma}{2\mu_n}\right)^2-\frac{\gamma c^2}{2g} \left(\left(\frac{k}{\mu_n}\right)^2+\left(\frac{\gamma}{2\mu_n}\right)^2 \right),\nonumber\\
			&\mathcal{B}(\mu_n,k) = \mu_nh\left[1- \left( \frac{\gamma}{2\mu_n}\right)^2 +\frac{\gamma c^2}{2g}\left(\left(\frac{k}{\mu_n}\right)^2+\left(\frac{\gamma}{2\mu_n}\right)^2-1\right) \right].\nonumber
		\end{align}
	\end{subequations}
	Consequently, the inner product $\langle\phi_n^\pm(\cdot,\cdot,k),\phi_p^\pm(\cdot,\cdot,\tilde{k})\rangle_{\mathcal{E}} $ is computed as
	\begin{subequations}
		\begin{align}
			\langle\phi_n^\pm, \phi_p^\pm\rangle_{\mathcal{E}} &=\int_{-\infty}^\infty\int_{-h}^0 \phi_n^\pm(x,z;k)\overline{\phi_p^\pm(x,z;\tilde{k})}\,{\rm d}z\,{\rm d}x +\frac{c^2}{g}\int_{-\infty}^\infty\phi_n^\pm(x,0;k) \overline{\phi_p^\pm(x,0;\tilde{k})}\,{\rm d}x,\nonumber\\
			&=\int_{-\infty}^\infty\langle f_n(z;k),f_p(z;\tilde{k})\rangle_h e^{\mp{\rm i}(k-\tilde{k}) x}\,{\rm d}x = 2\pi\frac{c^2}{g}\frac{\tilde{k}}{\tilde{\mu}_n\partial_k\tilde{\mu}_n}\delta(k-\tilde{k})\delta_{np}.
		\end{align}
	\end{subequations}
	Note that in the case of counter-propagating waves, we compute $\langle\phi_n^\pm(\cdot,\cdot,k),\phi_p^\mp(\cdot,\cdot,\tilde{k})\rangle_{\mathcal{E}}=0$, which is analogous to \eqref{eq:inner_product_opposite}.
	
\end{document}